\documentclass[11pt]{article}
\pdfoutput=1
\usepackage{latexsym}
\usepackage{amssymb}
\usepackage{epsf}
\usepackage{amsmath}
\usepackage{hyperref}
\usepackage{graphicx}
\usepackage{slashed}
\usepackage{color}
\usepackage{amsmath}

\makeatletter
\newcommand{\textarrow}[2][1]
  { \settowidth{\@tempdima}{#2}
    \stackrel{#2}
             {\makebox[#1\@tempdima][l]{\rightarrowfill}}
  }
\makeatother

\newcommand{\tr}{{\rm Tr}}

\newcommand{\Slash}[1]{{\ooalign{\hfil#1\hfil\crcr\raise.167ex\hbox{/}}}}

\newcommand{\beq}{\begin{equation}}  \newcommand{\eeq}{\end{equation}}
\newcommand{\bef}{\begin{figure}}  \newcommand{\eef}{\end{figure}}
\newcommand{\bec}{\begin{center}}  \newcommand{\eec}{\end{center}}
\newcommand{\non}{\nonumber}  
\newcommand{\laq}[1]{\label{eq:#1}}  

\newcommand{\Eq}[1]{Eq.~(\ref{eq:#1})}
\newcommand{\Eqs}[1]{Eqs.~(\ref{eq:#1})}
\newcommand{\eq}[1]{(\ref{eq:#1})}
\newcommand{\Sec}[1]{Sec.~\ref{chap:#1}}
\newcommand{\ab}[1]{\left|{#1}\right|}
\newcommand{\vev}[1]{ \left\langle {#1} \right\rangle }

\newcommand{\lac}[1]{\label{chap:#1}}
\newcommand{\SU}[1]{{\rm SU{#1} } }
\newcommand{\SO}[1]{{\rm SO{#1}} }
\def\({\left(}
\def\){\right)}

\def\diag{\mathop{\rm diag}\nolimits}

\def\O{\mathcal{O}}
\def\U{\mathop{\rm U}}

\def\ebq{\eeq \beq}

\newcommand{\AND}{~{\rm and}~}
\newcommand{\EV}{ ~{\rm  eV} }

\newcommand{\MEV}{ ~{\rm MeV} }
\newcommand{\GEV}{ ~{\rm GeV} }

\def\o{\over}
\def\a{\alpha}
\def\b{\beta}

\def\d{\delta}
\def\e{\epsilon}
\def\f{\phi}
\def\g{\gamma}

\def\k{\kappa}

\def\m{\mu}
\def\n{\nu}

\def\r{\rho}

\def\t{\tau}

\def\x{\xi}

\def\D{\Delta}
\def\G{\Gamma}

\def\ol{\overline}
\def\tl{\tilde}
\def\*{\dagger}

\begin{document}
\pagestyle{empty}

\begin{flushright}
KEK-TH-2064\\
\end{flushright}

\vspace{3cm}

\begin{center}

{\bf\LARGE Leptogenesis via Neutrino Oscillation Magic}  
\\

\vspace*{1.5cm}
{\large 
Yuta Hamada$^{1,2}$, Ryuichiro Kitano$^{1,3}$, and Wen Yin$^4$
} \\
\vspace*{0.5cm}

{\it
$^1$KEK Theory Center, Tsukuba 305-0801, Japan\\
$^2$Department of Physics, University of Wisconsin, Madison, WI 53706, USA\\
$^3$Department of Particle and Nuclear Physics\\
The Graduate University for Advanced Studies (Sokendai)\\
Tsukuba 305-0801, Japan\\
$^4$Institute of High Energy Physics, Chinese Academy of Sciences,
 Beijing 100049, China
}

\end{center}

\vspace*{1.0cm}

\begin{abstract}
{\normalsize 
The possibility of generating the baryon asymmetry of the Universe via
 flavor oscillation in the early Universe is discussed.
After the inflation, leptons are born in some states, travel in the medium, and are eventually projected onto flavor eigenstates due to the scattering via the Yukawa interactions. By using the Lagrangian of the Standard Model with the Majorana neutrino mass terms, $llHH$, we
 follow the time evolution of the density matrices of the leptons in
 this very first stage of the Universe and show that the CP violation in
 the flavor oscillation can explain the baryon asymmetry of the
 Universe.
 In the scenario where the reheating is caused by the decay of the
 inflaton into the Higgs bosons, the baryon asymmetry is generated by
 the CP phases in the Pontecorvo-Maki-Nakagawa-Sakata matrix and thus
 can be tested by the low energy neutrino experiments.
}
\end{abstract} 

\newpage
\baselineskip=18pt
\setcounter{page}{2}
\pagestyle{plain}
\baselineskip=18pt
\pagestyle{plain}

\setcounter{footnote}{0}

\section{Introduction}
The origin of the baryon asymmetry is a long-standing puzzle in the Standard Model and the Standard Cosmology. 
The successful generation of the baryon asymmetry requires the deviation
from the thermal equilibrium~\cite{Sakharov:1967dj}. 
In the standard inflationary cosmology with the Standard Model particles, there are two processes away from equilibrium: the electroweak phase transition and thermalization. The first process is related to the scenario of the electroweak baryogenesis~\cite{Kuzmin:1985mm} although some extension of the model is necessary to explain the amount of asymmetry. 
The thermalization era is also an ideal circumstance for the baryogenesis
since by definition the Universe is not in the thermal equilibrium. 
In general, after the inflation the Universe has experienced a thermalization era called reheating. The
baryogenesis at this stage is investigated recently~\cite{Hamada:2015xva, Takahashi:2015ula, Hamada:2016npz, Hamada:2016oft}.~(See also Refs.~\cite{Pascoli:2016gkf, Son:2018avk} for related recent works.)
There are also possibilities of having new particles in addition to the ones in the Standard Model, and CP-violating non-equilibrium decays of such particles generate baryon asymmetry such as in the scenario of thermal leptogenesis~\cite{Fukugita:1986hr}. 

The flavor oscillation during the reheating era plays the important role for baryogenesis. 
The flavor oscillation of the leptons can happen in the early Universe
when the high energy leptons from the decays of the inflatons go through the
medium. If the inflaton directly produces the left-handed leptons via
its decay, the initial quantum state is some vector in the flavor space,
that is generally not an eigenstate of the Hamiltonian in the
medium. The lepton flavors will later be ``observed'' by some
interactions. It has been discussed that the CP violation in the
quantum oscillation phenomena during this process can explain the baryon
asymmetry of the Universe~\cite{Hamada:2016oft}. 

In this paper, we describe the thermalization process with taking into account the quantum effects of the flavor oscillation.
To this end, we should employ the formulation in terms of density matrices rather than the classical Boltzmann equation. By solving the kinetic equation, we show that the baryon asymmetry of the Universe can be created within the Standard Model with the Majorana neutrino mass term, $llHH$, which explains the neutrino oscillation experiments. (For recent results, see Refs.~\cite{Abe:2017aap,  Abe:2017vif, NOvA:2018gge,  Gando:2013nba, An:2016ses}.) The baryon asymmetry is generated after a quite non-trivial evolution of the density matrices. The shape of the matrices changes during the travel in the medium and eventually settle into the diagonal form in the flavor basis by the lepton Yukawa interactions at later time. The CP-violating interactions in the neutrino sector create the difference between the lepton and anti-lepton density matrices during this evolution. 
We investigate scenarios where the leptons, $l$, are generated by the direct decays of inflaton, $\f$, or through the scattering of the Higgs bosons, $H$.  
We find that enough amount of asymmetry can be produced when the reheating
temperature of the Universe is beyond $10^8\GEV$ (lepton) or
$10^{11\text{--}12}$~GeV (Higgs). 
In particular, in the case where the
inflaton mainly decays into the Higgs boson 
(for example, through $\phi |H|^2$ interactions, see App.~\ref{app:inflaton_decay}.),
the CP phase stems from the Pontecorvo-Maki-Nakagawa-Sakata (PMNS)
matrix~\cite{Pontecorvo:1967fh,Maki:1962mu} and thus the scenario can be
tested in principle by future measurements.  The leptons produced by the
scattering through the $llHH$ operator undergo the flavor oscillations
in the medium, and the flavor dependent lepton asymmetry can be
generated by the CP violation in the oscillation. 
The flavor dependent asymmetries are converted to the net asymmetry via the $llHH$ interactions. The $llHH$ interactions are in fact more important than the gauge interactions at temperatures higher than $10^{14}$~GeV. For such high reheating temperatures, the first thermal bath is formed by the $llHH$ interactions, and many of Standard Model particles are out of thermal equilibrium. As temperature drops, the gauge and Yukawa interactions get gradually important. The evolution of the lepton density matrices during this thermalization era experiences flavor oscillations and multiple scattering processes before they settle into flavor diagonal forms. The CP-violating effects in the evolution explain the baryon asymmetry. In this high-temperature regime, the final baryon asymmetry does not depend on the detail of the inflaton properties, such as mass, branching ratio or the reheating temperature, as they are generated by the history of the medium.  The excess of the baryons above anti-baryons 
indicates that a combination of Dirac and Majorana phases is constrained to be in a certain range depending on the neutrino mass hierarchy
and the range of reheating temperature.
Interestingly, there exist
implications on the effective Majorana neutrino mass, $m_{\nu ee}$, which
determines the rate of the neutrino-less double beta decay.

This paper is organized as follows. In Sec.~\ref{sec:quantum_equation},
we briefly summarize the quantum equation describing the flavor
oscillation. The setup of our scenario is described in
Sec.~\ref{sec:basic_scenario}. Our kinetic equation describing the time
evolution of the density matrix is presented in
Sec.~\ref{sec:kinetic_equation}, and it is solved numerically in
Sec.~\ref{sec:numerical_result}. The explanation of the behavior of the
solution based on the analytic calculation is given in
Sec.~\ref{sec:analytical_understanding}. 
The last section is devoted to conclusions.

\section{Quantum equation for lepton flavor}\label{sec:quantum_equation}

The lepton oscillation and its effect on the lepton asymmetry can be
described by the time evolution of the density matrix in the flavor
space~\cite{Sigl:1992fn}. By using the free Hamiltonian of the
left-handed leptons, $l_i$,
\begin{align}
 H_0& = \int {d^3{\bf p} \over (2 \pi)^3 }
\left(
a_i^\dagger ({\bf p}, t) \Omega_{ij} ({\bf p}) a_j ({\bf p}, t)+
b_j^\dagger ({\bf p}, t)\Omega_{ij} ({\bf p}) b_i ({\bf p}, t)
\right),
\end{align}
the evolution of the density matrices of $l_i$ and its anti-particle
defined by
\begin{align}
 \rho_{ij} ({\bf p}, t)& = \langle a_j^\dagger ({\bf p}, t) a_i ({\bf
 p}, t)\rangle / V, \quad
 \bar \rho_{ij} ({\bf p}, t) = \langle b^{\dagger}_{i} ({\bf p}, t) b_j ({\bf
 p}, t)\rangle / V,
\label{eq:density}
\end{align}
are given by
\begin{align}
 \dot \rho ({\bf p}, t) & = -i \left[
\Omega ({\bf p}), \rho ({\bf p}, t)
\right],
\quad
 \dot {\bar \rho} ({\bf p}, t)  = i \left[
\Omega ({\bf p}), \bar \rho ({\bf p}, t)
\right].
\label{eq:rhoeq}
\end{align}
Here $V$ is the volume of the system, $V = (2 \pi)^3 \delta^3 (0)$.  
The expectation values in \Eq{density} are taken by the state to describe the Universe.
The flavor indices, $i$ and $j$, are ordered differently for particles and anti-particles
in Eq.~\eqref{eq:density}. In this definition, $\rho$ and $\bar \rho$
transform, respectively, as $U \rho U^\dagger$ and $U \bar \rho U^\dagger$
under a unitary rotation of flavors. The density matrix of the lepton
asymmetry is naturally defined by
\begin{align}
 \Delta_{ij} ({\bf p})& = \rho_{ij} ({\bf p}) - \bar \rho_{ij} ({\bf p}).
\end{align}
Due to the notation in Eq.~\eqref{eq:density}, CP invariance indicates
$\rho_{ij}=\bar{\rho}_{ji}$. Even if CP is conserved, the off-diagonal
components of $\Delta_{ij}$ can be nonzero while the diagonal entries
should vanish.

The trace of the matrix, $\Delta$, describes the total asymmetry for left-handed leptons (right-handed lepton should be added to get total one) which is independent of the basis. Note that $\rho$ and $\bar \rho$ evolve
differently as in Eq.~\eqref{eq:rhoeq}. The difference serves as the
``strong phase'' in the CP violation in the flavor oscillation. The
asymmetry evolves as
\begin{align}
 \dot \Delta ({\bf p})& = -i \left[
\Omega ({\bf p}), \rho({\bf p}) + \bar \rho ({\bf p})
\right].
\label{eq:Deltaeq}
\end{align}
Even if $\Delta ({\bf p}) = 0$ at some time, the non-trivial matrix can
be generated at a later time if $\rho + \bar \rho$ does not
commute with the Hamiltonian, while the trace is kept vanishing.

The effects of the interaction term in the Hamiltonian, $H_{\rm int}$,
have been discussed in Ref.~\cite{Sigl:1992fn}. By using the perturbative expansion and
the approximation of the instantaneous collisions, the evolution at a
time $t=0$ is
given by
\begin{align}
 \dot \rho ({\bf p})& = 
- i \left[
\Omega ({\bf p}), \rho ({\bf p})
\right]
+ i \Big \langle \left[
H_{\rm int}^0 (0), 
a_j^\dagger ({\bf p})
a_i ({\bf p}) / V
\right] \Big \rangle
\nonumber \\
&
-{1 \over 2} \int_{-\infty}^\infty dt
\Big \langle \left[
H_{\rm int}^0 (t), 
\left[
H_{\rm int}^0 (0), 
a_j^\dagger ({\bf p})
a_i ({\bf p}) / V
\right]
\right] \Big \rangle,
\label{eq:master}
\end{align}
where $H_{\rm int}^0 (t)$ is the interaction Hamiltonian in the
interaction picture, $e^{i H^0 t} H_{\rm int} (t=0) e^{-i H^0 t}$. One
can use this equation for any time $t$ by treating each collision
independently.

We use the above formulation for the discussion of the flavor
oscillation of the leptons from the inflaton decay.

\section{Basic scenario}\label{sec:basic_scenario}
As the simplest example for the mechanism, we consider the Standard
Model with the Majorana masses for neutrinos:
\begin{align}
 {\cal L} & = {\cal L}_{\rm SM} 
 - {\kappa_{ij} \over 2} (\bar l_i^c P_L l_j) H H + {\rm h.c.}
\end{align}
The indices of $\SU(2)_L$ gauge interaction are implicit. One can obtain
the model, for example, by integrating out right-handed
neutrinos~\cite{Minkowski:1977sc,Yanagida:1979as,GellMann:1980vs,Glashow:1979nm,Mohapatra:1979ia}.
We assume that right-handed neutrinos (or any other alternative) are
sufficiently heavy, and do not show up in the history of the Universe.
The Lagrangian of the Standard Model $({\cal L}_{\rm SM})$ contains the
Yukawa interaction of the leptons: 
\begin{align}
 {\cal L}_{\rm Yukawa}& =  -y_i \bar l_i H P_R e_i + {\rm h.c.,}
\end{align}
where we take the basis where $y_i$ is real and positive.  In this
basis, the symmetric matrix $\kappa$ is given by 
\begin{align}
 \kappa \langle H \rangle^2 & = U_{\rm PMNS}^* m_\nu U_{\rm PMNS}^\dagger,
\end{align}
where $m_\nu=\diag{(m_{\nu1},m_{\nu2},m_{\nu3})}$ is the real, non-negative and diagonal matrix of the neutrino
masses and $\langle H \rangle \simeq 174$~GeV. 
There are three CP phases in the PMNS matrix~\cite{Patrignani:2016xqp}:
a Dirac phase $\d$, and two Majorana phases $\a_M=\a_{21}$ and
$\a_{M2}=\a_{31}$. If the lightest neutrino is massless, we will take
the redundant parameter $\a_{M2}=0$. Throughout this paper, we use the
indices $(i, j, ... )$ for the flavor basis while the indices $(\alpha,
\beta, ... )$ for the mass basis. Namely, $i, \ j=e, \ \mu,\ \tau$ and
$\alpha, \ \beta=1,\ 2,\ 3$.

Let us mention the validity of the effective theory in terms of the $llHH$ interaction.
The perturbative expansion makes sense when the typical energy, $E$, of the scattering process, {e.g.} the temperature, satisfies
\beq
\laq{pert}
 {\max{[m_{\nu}]}  \over 16\pi^2  \vev{H}^2}E \lesssim 1.
\eeq
When there are new particles, such as the right-handed neutrinos, our treatment based on the effective theory is not accurate when the reheating temperature goes beyond the mass scale of such particles. Although we do not study those scenarios, the same mechanism we describe below may still work even in such cases.

We assume that, after the inflation, the decay of inflaton reheats the
Universe and the left-handed leptons are produced as daughter particles.
For example, if the production is directly from the inflaton decay,
the lepton state is parametrized as
\begin{align}
 | l_\phi \rangle & = V_i | l_i \rangle.
\label{eq:lphi}
\end{align}
The coefficient $V_i$ is a normalized vector. In this setup, the
inflaton sector is characterized by the reheating temperature $T_R$, the
vector $V_i$ and the branching fraction to left-handed leptons $B$.
Even if the inflaton does not directly decay into leptons but decays
into Higgs bosons,  
the leptons are, in turn, generated by the scattering of the high-energy Higgs bosons with the medium through the Yukawa or
$llHH$ interactions, and hence the parameterization above is still
useful.  Note that from the constraint on the tensor-to-scalar ratio in
the curvature fluctuations, $r<r_{\rm max}\simeq
0.12$~\cite{Ade:2015xua}, the reheating temperature is bounded from
above:
\beq 
T_R\lesssim 10^{16}\GEV \times{\( {g_*(T_R)\over 100}\)^{-1/4}} 
\({r_{\rm max} \over 0.12}\)^{1/4}. 
\laq{tts}
\eeq
We also assume that the time scale for the thermalization is much faster
than the decay rate of the inflaton, $\Gamma_\phi$. In that case,
$\Gamma_\phi \sim T_R^2 / M_{\rm P}$. (This assumption will be
justified later.) The thermalized component of the radiation with
temperature $T_R$ is quickly produced during the reheating era.
Under this assumption, we follow the time evolution
of the density matrices while the thermal plasma with temperature $T_R$
already exists as the initial condition.

By the interaction with the thermal plasma, the leptons
 quickly lose their energies by scattering processes, and
eventually they are annihilated by hitting their anti-particles. In the course of
this non-equilibrium process, the leptons undergo the flavor oscillation
since the thermal masses of the leptons are flavor dependent. 
Working in the ``mass'' basis, where the neutrino mass matrix is
diagonal in the vacuum, the generation (in the mass basis) dependent
lepton numbers are produced via the CP violation in the oscillation,
while net asymmetry is not created.
These flavor dependent lepton asymmetries are partially washed out by
the scattering via $llHH$ terms. Since the rate of this process is
generation dependent, the net asymmetry is produced. 
Depending on the decay modes and the reheating temperatures, there are other scenarios 
which generate flavor or chirality dependent lepton asymmetries. We will discuss each scenario in detail in Sec. \ref{sec:analytical_understanding}.

The ingredients of this baryogenesis are the Yukawa interactions and the
$llHH$ interactions, both of which are measurable at low energy. If both
of them are in the thermal equilibrium, one cannot obtain the baryon
asymmetry. Rather, any baryon or lepton asymmetry would be erased. The
important fact is that when $llHH$ interaction is effective at high
temperatures, the Yukawa interaction is ineffective due to the cosmic
expansion. The opposite is true at low temperatures. Therefore, as we
will see later, baryon asymmetry can be generated in a wide range of
reheating temperatures.

\section{Kinetic equation}\label{sec:kinetic_equation}

We perform a numerical computation of the lepton asymmetry by solving
the kinetic equations for the density matrices of leptons and anti-leptons.
The oscillation, decoherence by scattering, annihilation, creation and the lepton
number generation are described by the equations.

Following the formalism in Sec.~\ref{sec:quantum_equation}, the kinetic
equation used in this paper is presented below. The starting point is the
master equation obtained from Eq.~\eqref{eq:master}. The kinetic equation is
summarized in the form of
\beq
\label{eqnu}
  i\frac{d \rho ({\bf p})}{dt} = 
[\Omega ( {\bf p} ) , \rho ( {\bf p} )] - 
  \frac{i}{2} \{ \Gamma_{ {\bf p}}^d, \rho ( {\bf p} )  \} +
  \frac{i}{2} \{ \Gamma_{ {\bf p}}^p, 1-\rho ( {\bf p} )  \} , 
\eeq
where flavor indices are implicit~\cite{Sigl:1992fn}.  The first term describes the
oscillation while the second and third terms correspond to the
destruction and production processes, respectively.  The Hamiltonian
$\Omega ({\bf p})$ can be parametrized as
\begin{align}
 \Omega_{ij} ({\bf p})& = |{\bf p}| \delta_{ij}+ \d \Omega_{ij} ({\bf p}),
\end{align}
where the thermal correction $\d \Omega ({\bf p})$ can be written as
\begin{align}
\d \Omega_{ij} ({\bf p})
\simeq 	
{y_i^2 T^2 \over 16 |{\bf p}|}\delta_{ij}+{0.046}(\kappa^* \kappa)_{ij}{T^4\over |{\bf p}|},	
& \quad \text{for $|{\bf p}|\gtrsim T$.}
\label{eq:thermalmass}
\end{align}
Here we do not include terms which are proportional to the unit matrix
since they do not contribute to the kinetic equation.
The coefficients are evaluated under the assumption that the left-handed leptons, right-handed leptons and the
Higgs bosons are all thermalized. The second term is evaluated by
calculating the two loop thermal diagram with $llHH$ interaction.  At a
high temperature where the Yukawa interactions are not effective, the
right-handed leptons are not in the thermal bath and the coefficient of
the first term is modified. For simplicity, we do not consider the effects of the change of the coefficient in the numerical analyses.

We focus on the following two components of the density matrices:
\beq
\left(\r_{\bf k}\right)_{ij}=
\int_{|{\bf p}| \sim |{\bf k}|} {d^3 {\bf p}\over (2 \pi)^3}
\, 
{\r_{ij}({\bf p},t) \over s },
\ebq 
\left(\d \r_T\right)_{ij}=\int_{\ab{{\bf p}}\sim T}{{d^3 {\bf p}\over (2
\pi)^3}
\(
{\r_{ij}({\bf p}) \over s}
-
{\r^{\rm eq}_{ij}({\bf p}) \over s}
\) 
},
\eeq
and those for anti-leptons. Here $s$ is the entropy density. 
The first component, $\rho_{\bf k}$, represents the high energy leptons
produced by the inflaton decay with initial typical momentum,
\begin{align}
 |{\bf k}| =
m_\phi 
\left( {t_R \over t} \right)^{1/2},
\end{align}
where 
\begin{align}
\laq{tR}
 &t_{R}= \left({g_* \pi^2\over
30} \cdot { T_R^4 \over 3M_{\rm P}^2 }\right)^{-1/2},
\end{align}
is the time at the inflaton decay.  $g_*$ is the effective degree of freedom for the thermal plasma.  The high energy leptons lose their
energies by redshift and scattering processes.
The second component, $\delta \rho_T$, represents leptons
with the typical momentum $|{\bf p}| \sim T$, with the temperature
$T\sim T_R({t_{R} / t})^{1/2}$.
Here $\r_{ij}^{\rm eq}={ \d_{ij} / ( e^{|{\bf p}|/T}+1} )$ represents
the density matrix in the thermal equilibrium.

In terms of $\r_{\bf k}$ and $\d \r_T$, the kinetic equation becomes
\begin{align}
  i\frac{d \rho_{ {\bf k}}}{dt} = [\Omega_{ {\bf k}} , \rho_{ {\bf k}}] - 
  \frac{i}{2} \{ \Gamma_{ {\bf k}}^d, \rho_{ {\bf k}}  \},
\label{eq:kinK}
\end{align}
\begin{align}
  i\frac{d \d \rho_T}{dt} = [\Omega_{ T} , \d \rho_{ T}] - 
  \frac{i}{2} \{ \Gamma_{ T}^d, \d \rho_{ T}  \}+{i} \d \G_{T}^p ,
\label{eq:kinT}
\end{align}
where $\Omega_{\bf k} = \Omega ({\bf |p|} = {\bf |k|})$ and $\Omega_T = \Omega
({\bf |p|}=T)$.
The destruction and production rates for leptons are given by
\begin{align}
\left(\Gamma_{ {\bf k}}^d\right)_{ij}
	&\simeq
C \a_2^2  T \sqrt{\dfrac{T}{|{\bf k}|}} \d_{ij}
+ \dfrac{9 y_t^2}{64 \pi^3 |{\bf k}| }T^2 (\d_{i\tau}\d_{\tau j} y_\tau^2
 +\d_{i\mu}\d_{\mu j} y_\m^2 )
+ \dfrac{21 \zeta{(3)}}{32\pi^3} (\k^* \cdot \k )_{ij}{T^3},
\label{eq:GammaK}
\end{align}
\begin{align}
\left(\Gamma_{ T }^d\right)_{ij} \simeq
C' \a_2^2T \d_{ij}
+\dfrac{9 y_t^2}{ 64 \pi^3  }T (\d_{i\tau}\d_{\tau j}y_\tau^2+\d_{i\mu}\d_{\mu j}y_\m^2) 
+ \dfrac{21\zeta{(3)}}{32 \pi^3} (\k^* \cdot \k )_{ij}{T^3} ,
\label{eq:GammaT}
\end{align}
\begin{align}
\left(\d  \G_T^{p}\right)_{ij}
	&\simeq
C \a_2^2  T \sqrt{\frac{T}{|{\bf k}|}} \left(\r_{\bf k}\right)_{ij}
-C' \a_2^2T\left(\d \ol{\r}_T
\right)_{ij} \nonumber\\
& 		
+ {3\zeta{(3)}\o 8 \pi^3} (\k^* \cdot \( \ol{\r}_{\bf k} -3/4
 {\r}_{\bf k} \)^t\cdot \k )_{ij}{T^3}+ {3\zeta{(3)}\o 8\pi^3} (\k^*
 \cdot \(\d \ol{\r}_T-3/4\d\r_T\)^t \cdot \k )_{ij}{T^3}.
\label{eq:deltaGamma}
\end{align}
Here 
the superscript $t$ denotes the
transpose of the matrix.  
The equations for the anti-leptons can be obtained by exchanging $ \rho$
by $\bar \rho$ everywhere while changing the sign of $\Omega$'s.
The solution of the equations at $t \to \infty$ is $\rho_{\bf k} = \bar
\rho_{\bf k} = \delta \rho_T = \delta \bar \rho_T = 0$ when we ignore
the expansion of the Universe as is always the case. The expansion of
the Universe makes the $llHH$ interaction ineffective at later time,
leaving non-vanishing lepton asymmetry as we discuss below.

In the above equations, we have used the densities of the Higgs boson
and the right-handed leptons as the one in the thermal equilibrium.
In the actual numerical computation, the kinetic equations of
right-handed leptons are taken into
account~(c.f.~Ref.\cite{Asaka:2011wq} and App.\ref{App:right-handed_leptons}).
The effects of the U(1)$_Y$ gauge interactions are also included.

The first terms in Eqs.~\eqref{eq:GammaK} and \eqref{eq:deltaGamma}
denote the thermalization process through the $\SU(2)_L$ gauge interactions
where the Landau-Pomeranchuk-Migdal (LPM)
effects~\cite{Landau:1953um,Migdal:1956tc} are taken into account.  The
coefficient $C=\O(1)$ represents the theoretical uncertainties in the
rates and also in the energy distributions of the inflaton decay
product. 
This term converts the high energy leptons into low energy ones while
the matrix structure is untouched.

The second terms in Eqs.~\eqref{eq:GammaK} and \eqref{eq:GammaT}
describes the scattering via the Yukawa interactions. These terms, if
strong, bring the density matrices into the diagonal form in the flavor
basis and thus prevent the oscillation phenomena. (Similar formula can
be found in Refs.~\cite{Akhmedov:1998qx, Abada:2006fw, Asaka:2011wq}.)\footnote{The 2 to 2 scattering with gauge boson process and 1 to 2 (inverse) decay process with LPM effect may contribute to the Yukawa interaction rate and effectively alter the overall factor. (c.f. Refs.\cite{Besak:2012qm, Ghiglieri:2017gjz}.) However, this would not change our result significantly.}

The terms with the coefficient $C'$ in Eqs.~\eqref{eq:GammaT} and
\eqref{eq:deltaGamma} represent the pair annihilation and creation of
leptons, respectively. 
This process is important for low energy leptons $({\bf p} \sim T)$. For high-energy leptons, the rates are suppressed by the Boltzmann factor or $T/m_\phi$. 
A precise estimation of the rate requires the
inclusions of the infrared regularization as well as the LPM
effects~\cite{Arnold:2001ms}. We put a parameter $C' = {\cal O}(1)$ which
represents the theoretical uncertainty.  This term brings the total
density matrix $\rho + \bar \rho$ to a one proportional to the unit
matrix. By Eq.~\eqref{eq:Deltaeq}, the flavor oscillation stops when
this interaction becomes important as expected.

Finally, the terms with $\kappa$ are the effects of the scattering via
$llHH$ interactions. These terms become unimportant at low
temperatures. This interaction brings the density matrices into the
diagonal form in the mass basis and lets the asymmetries flow towards zero.

Before ending this section, let us give the kinetic equations for the trace of the asymmetry matrices, 
\begin{align}
 \tilde \Delta_{\bf k}& = \rho_{\bf k} - \bar \rho_{\bf k}, \quad
 \tilde \Delta_{T} = \delta \rho_T - \delta \bar \rho_T,
\end{align}
for later convenience.
That is 
\begin{align}
 \frac{d }{dt} \tr{[\tl{\D}_{ {\bf k}}+\tl{\D}_{ {T}}]}&=- {21 \zeta (3) T^3 \over 16 \pi^3 \langle H \rangle^4} \tr{[(\tl{\D}_{ {\bf k}}+\tl{\D}_{ {T}})m_\n^2]} \non \\
 &-\frac{9 y_t^2 T^2}{64 \pi^3 |{\bf k}| } \( y_\t^2(\tl{\D}_{ {\bf k}})_{\t\t}+y_\m^2(\tl{\D}_{ {\bf k}})_{\m\m}\)-\frac{9 y_t^2T}{64 \pi^3 } \( y_\t^2(\tl{\D}_{ {T}})_{\t\t}+y_\m^2(\tl{\D}_{ {T}})_{\m\m}\)+\cdots.
 \label{eq:trace}
 \end{align}
The terms in the first and second lows are essentially different.  The
first term, i.e. the wash-out term, decreases or increases the asymmetry for the
left-handed leptons, while the terms in the second row transfer the
asymmetry into the right-handed leptons without changing the net lepton asymmetry.

\section{Numerical results}\label{sec:numerical_result}

\begin{figure}[!t]
\begin{center}  
   \includegraphics[width=75mm]{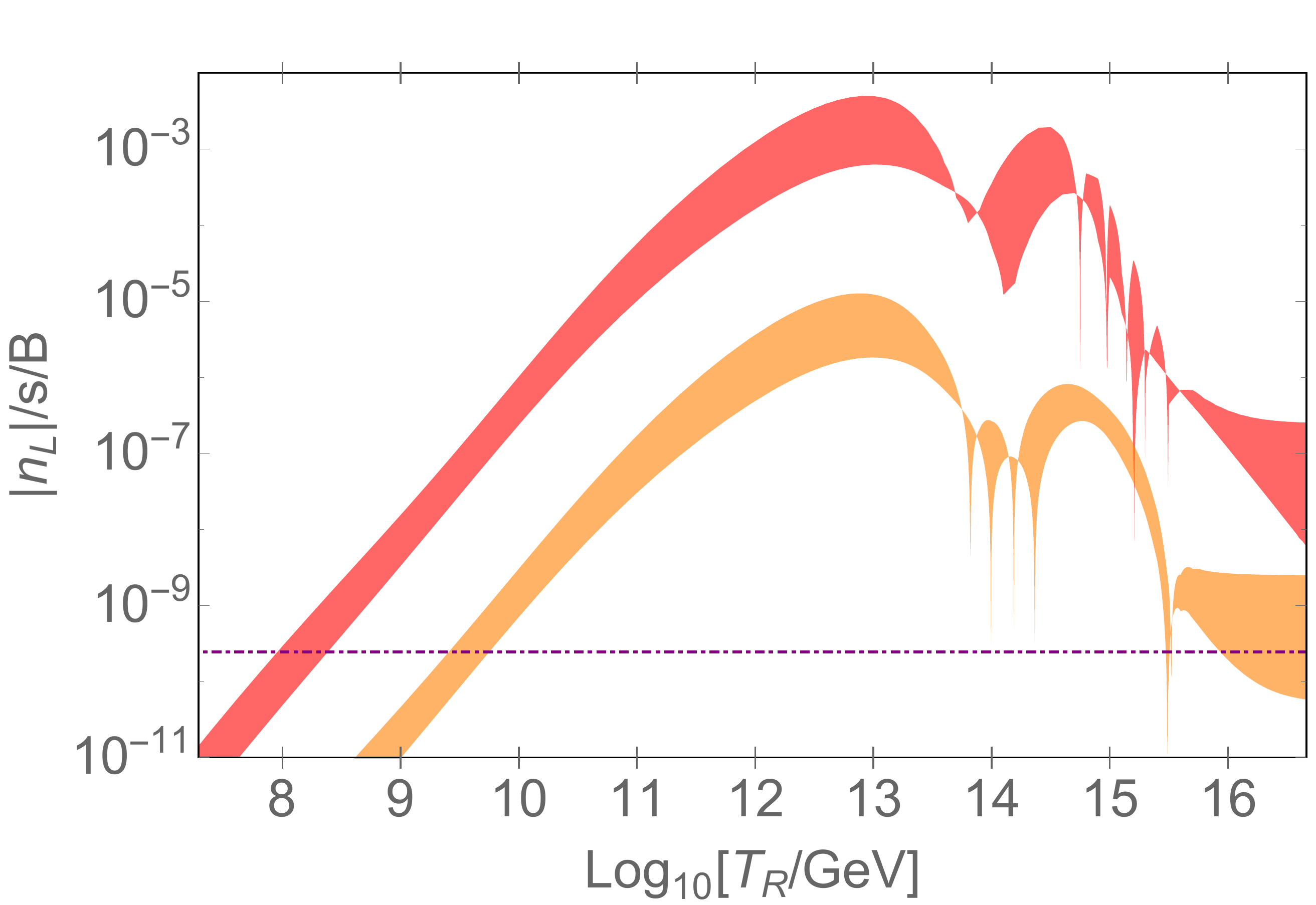}
      \includegraphics[width=75mm]{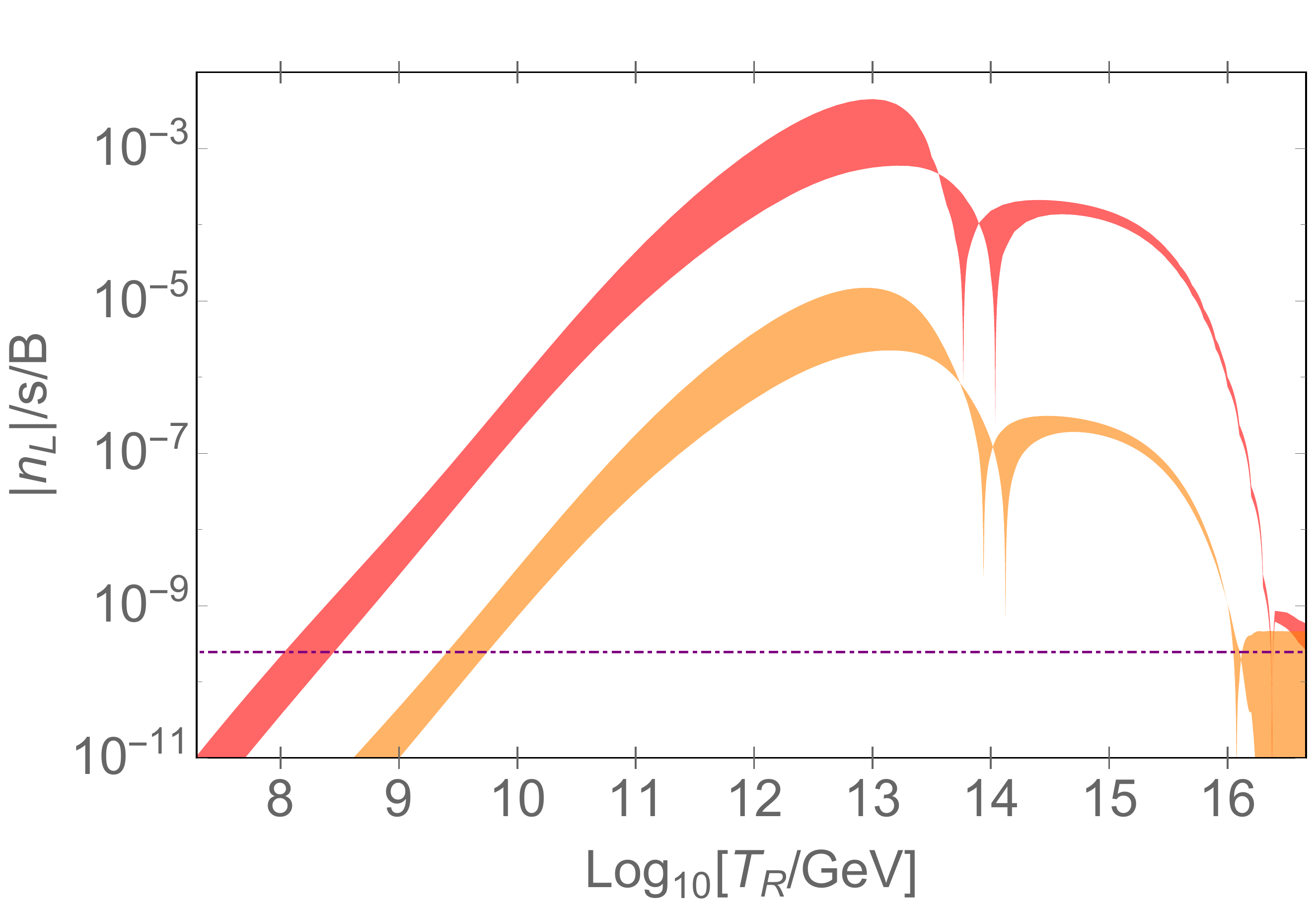}
      \end{center}
\caption{
The dependence of lepton asymmetry on the reheating temperature with
 $\a_M=0.3\pi, \d=-\pi/2$, $V={1\over\sqrt{3}}(1,1,1)$ with the normal (inverted) mass hierarchy with one massless neutrino, $ m_{\n {\rm lightest}}= 0\EV$, in the left (right) panel. The red and orange bounds represent $m_\f/T=1$ and $100$, respectively.
 Each band corresponds to the variance of $C=C'$ between $1/3$ and $3$. The purple line represents the required lepton asymmetry. The shaded region may be invalid as the effective theory calculation. }
\label{fig:Numerical_Result}
\end{figure}

The kinetic equations in Eqs.~\eqref{eq:kinK} and \eqref{eq:kinT} are
solved numerically by setting initial conditions which describe the
inflaton decay.
For the case where the inflaton decays into a single linear combination
of the flavor eigenstates as in Eq.~\eqref{eq:lphi}, the initial density
matrices are given by
\begin{align}
&\r_{\bf{k}}|_{t=t_R}=\ol{\r}_{\bf{k}}|_{t=t_R}= { {\cal N} V_i V^*_j },
\quad \d \r_T|_{t=t_R} = \d \bar \r_T|_{t=t_R}= 0.
\label{eq:initial}
\end{align}
The normalization factor ${\cal N}$ is given by
\begin{align}
{{\cal N}} 
= {3\over4}{T_R\over m_\phi} B,
\end{align}
where $B$ is the branching fraction of the inflaton into high-energy
leptons. 
For the case of direct decays into leptons, the initial distributions depend on the unspecified main decay mode. Taking  the thermal distribution as the initial condition provides a conservative estimate of the baryon asymmetry as we will see later. 
In general, the decay product can be a weighted sum of
different states, i.e., a mixed states, such as ${\sum_{a} {\cal N}_a
(V_i^a)^* V^a_j }.$ 
An interesting possibility is that the inflaton mainly decays into the
Higgs bosons. The high energy Higgs bosons, in turn, hit the leptons or
Higgs bosons in the medium and produce the high energy leptons through
the $llHH$ interactions. In that case, the density matrices are
given by
\begin{align}
\laq{Higgsdecay}
(\rho_{\bf k})_{ij} &=
(\bar \rho_{\bf k})_{ij} =
 {\cal N}
\cdot {{21\zeta(3)\over32\pi^3}(\kappa^*\kappa)_{ij} T_R^3\over C\alpha_2^2 T_R \sqrt{T_R\over m_\phi}}
\sim 7\times10^{-2}  \cdot 
{\cal N}
\left({m_\phi\over10^{15}\GEV}\right)^{1/2}
 \left({T_R\over10^{13}\GEV}\right)^{3/2}.
\end{align}
Here $\mathcal N$ is the same as the previous definition, but $B\sim \O(1)$ is the branching ratio of the inflaton decays to Higgs boson. 
See App.~\ref{App:parameters} for the parameters used in the calculation.
The factor can be thought of as the branching fraction of the Higgs
boson into leptons in the medium. The denominator represents the inverse
of the lifetime of the high energy Higgs boson in the medium. The
contributions from the Yukawa interactions are always negligible in the
temperature range of our interest although they are included in the
numerical calculations.
For a very high reheating temperature where the $llHH$ interaction is
stronger than the gauge interactions, the denominator should be replaced
by the trace of the numerator.

Only for the cases of the inflatons decay into the Higgs bosons, the
medium leptons are set to be zero initially, rather than assuming the
thermal distributions in Eq.~\eqref{eq:initial}:
\beq
\laq{inihig}
\quad \d \r_T|_{t=t_R} = \d \bar \r_T|_{t=t_R}= -0.004 \({100\o
g_{*s}(T_R)}\) \delta_{ij}.
\eeq
This deviation from the thermal distributions plays the relevant role
for high reheating temperatures, where the rates of the gauge
interactions are slower than the expansion rate of the Universe and thus
the thermalization process can be flavor dependent. By the strongest interactions in each temperature regime, the thermal components are created within the time scale $t_R$. The effects of taking \Eq{inihig} as initial condition are not important for $T_R \lesssim 10^{13}$~GeV. 
The temperature dependences of $g_*$ and $g_{*s}$ are not included in
the calculation. They depend on the detail of the thermalization
histories. In the numerical calculation, we take $g_{*s}=g_*=100$.

\begin{figure}[!t]
\begin{center}  
   \includegraphics[width=75mm]{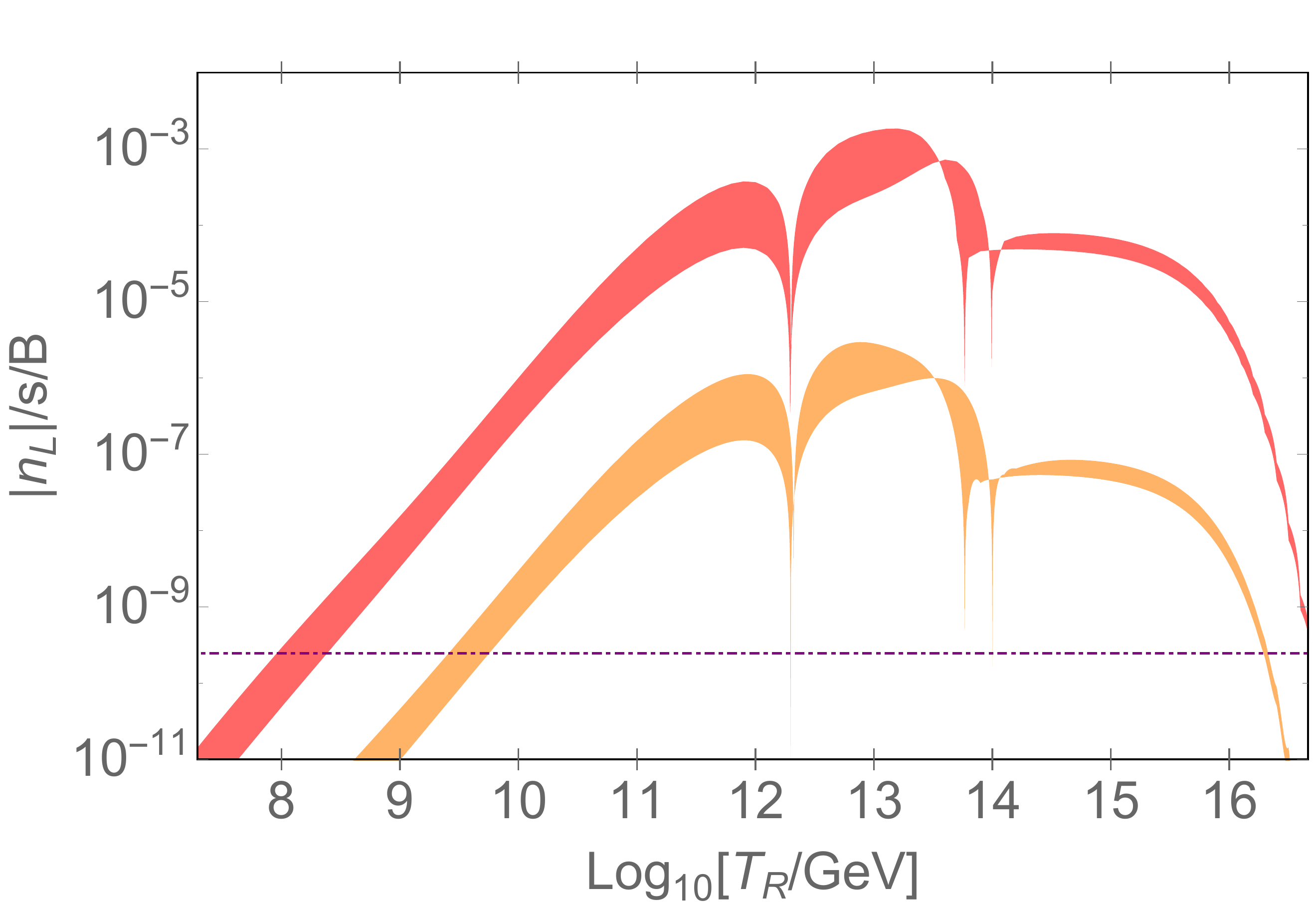}
  \includegraphics[width=75mm]{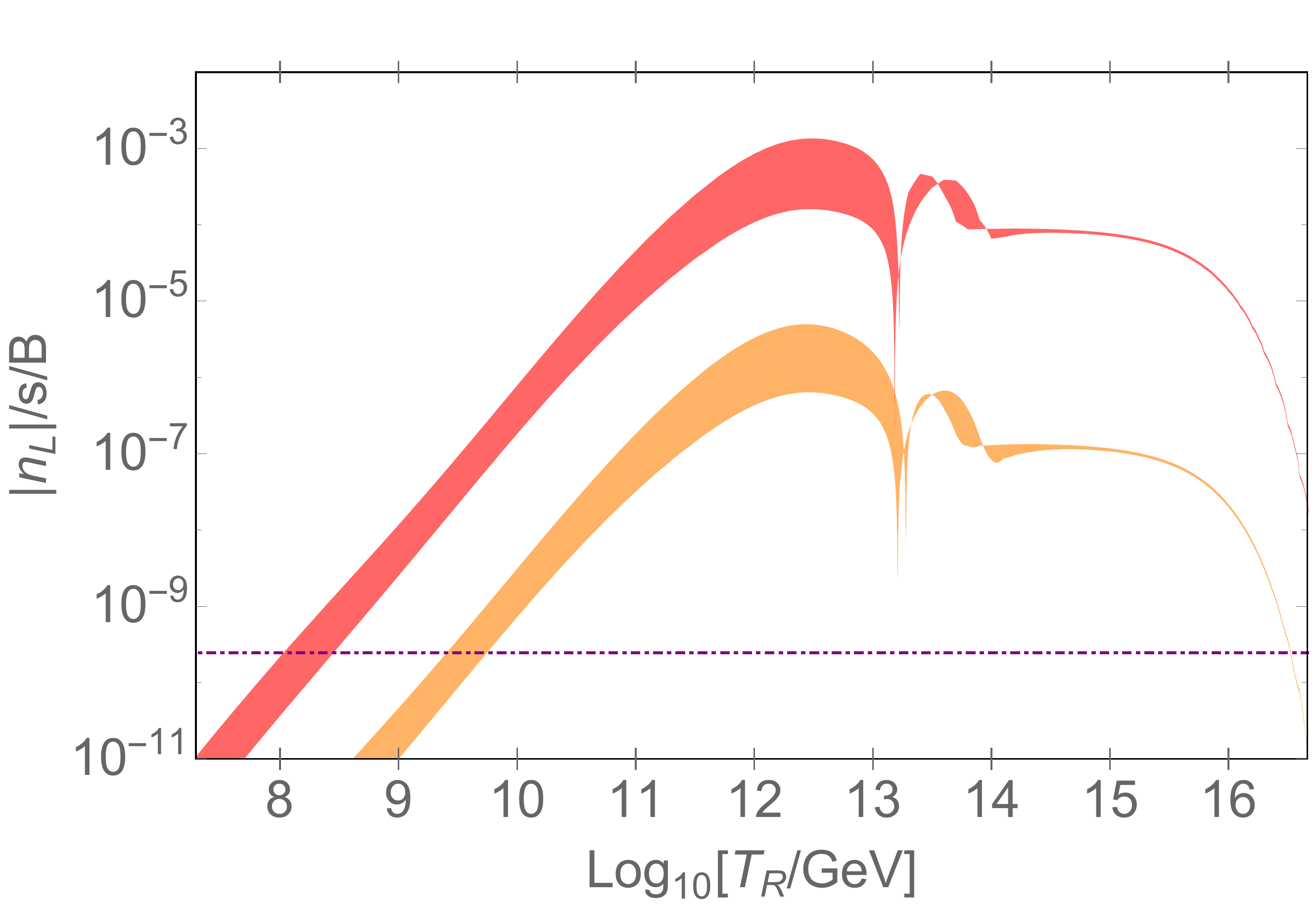}
\end{center}
\caption{
The dependence of lepton asymmetry on the reheating temperature with
degenerate neutrino mass, $m_{\n{\rm lightest}}= 0.07~\EV$. We take $\alpha_{M2}=0$. The neutrino masses are in normal (inverted) ordering in the left (right) panel.
The other parameters are the same as Fig.~\ref{fig:Numerical_Result}.
}\label{fig:Numerical_Result3}
\end{figure}

The kinetic equations are solved numerically
and 
\begin{align}
{n_L\over s}={{\rm Tr}\left(
\tilde \D_{\bf k}+
\tilde \Delta_T
+\tilde \D_R\right)},
\end{align}
is evaluated at a low enough temperature, where $n_L/s$
is already frozen to a constant. Here ${\tilde \D_R}$ is the asymmetry
transferred into the right-handed leptons through the Yukawa
interaction.

The baryon asymmetry of the Universe is measured to be~\cite{Ade:2015xua}
\begin{align}
{n_B\over s}=(8.67\pm0.05)\times10^{-11}.
\end{align}
By assuming that the asymmetry is created above the electroweak scale, the
chemical equilibrium of the sphaleron
process~\cite{Klinkhamer:1984di,Kuzmin:1985mm} tells us
\begin{align}
\laq{nL}
{n_L\over s}\simeq -{79\over28}{n_B\over s}=-(2.45\pm0.01)\times10^{-10}.
\end{align}

We show in Fig.~\ref{fig:Numerical_Result} the absolute value of the
lepton asymmetry by varying $T_R$ while fixing the ratio $m_\phi/T_R=1$
(red) or 100 (orange) with $\d=-\pi/2,\a_M=0.3\pi$. The vector $V$ is set to
be $V \propto (1,1,1)$.
The left and right panels, respectively, correspond to the normal and
inverted hierarchies of neutrino masses.
The lightest neutrino mass, $m_{\n {\rm lightest}}$, is set to be zero
for both cases. 
The bands represent the uncertainties from the $C$ and $C'$ factors in
Eqs.~\eqref{eq:GammaK}, \eqref{eq:GammaT}, and \eqref{eq:deltaGamma}. 
We took $C = C'$ and varied the value from $1/3$ to $3$.
The same figures for degenerate neutrino cases are shown in
Fig.~\ref{fig:Numerical_Result3} where the lightest neutrino mass is
taken to be $m_{\nu {\rm lightest}} = 0.07$~eV, and $\alpha_{M2}=0$. 
We note that from the condition \eq{pert} by taking $E\lesssim \sqrt{m_\phi T_R}$, which is the typical energy of scattering between the high-energy leptons and the ambient thermal plasma, we get
\beq
T_R \lesssim 10^{16}\GEV  \sqrt{ 100 \over m_\phi/T_R }  \({ 0.05 \EV \over \max{[m_{\nu}]}} \).
\eeq
This is around the bound of \Eq{tts}.

One can see that the baryon asymmetry can be explained for $T_R \gtrsim10^8\GEV$.\footnote{For such low reheating temperatures the Hubble parameter during inflation can be as low as $H_{\rm inf}=
\O(10)\MEV$.  Recently, it was shown that with such a low-scale inflation
the QCD axion with decay constant around GUT or string scale can be the dark
matter \cite{Graham:2018jyp,Guth:2018hsa}.}  We stress that the contribution discussed here always exists in any models to explain the neutrino masses by the effective $llHH$ terms. Notice that we have
assumed the perturbative decay of the inflaton and thus the reheating
temperature is taken to be below the mass of the inflaton. However, a
non-perturbative reheating allows the temperature to be much higher than the inflaton mass, which may further enhance the
asymmetry. (See c.f. Refs.~\cite{Daido:2017wwb,Daido:2017tbr} for enhancing efficiency for the conversion of the inflaton energy.) We leave the analysis of lepton asymmetry in this case for the future study.

\begin{figure}[!t]
\begin{center}  
   \includegraphics[width=75mm]{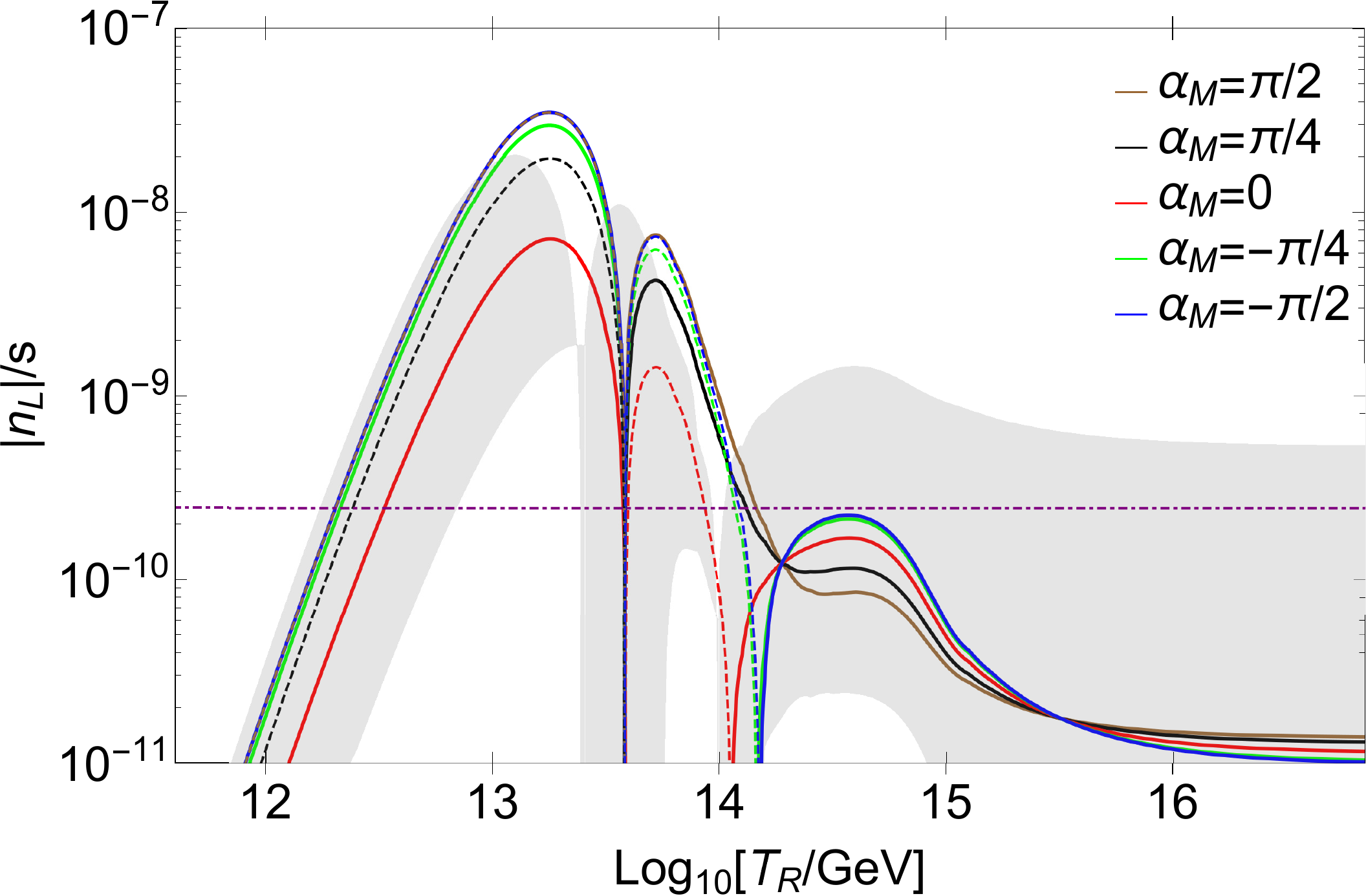}
      \includegraphics[width=75mm]{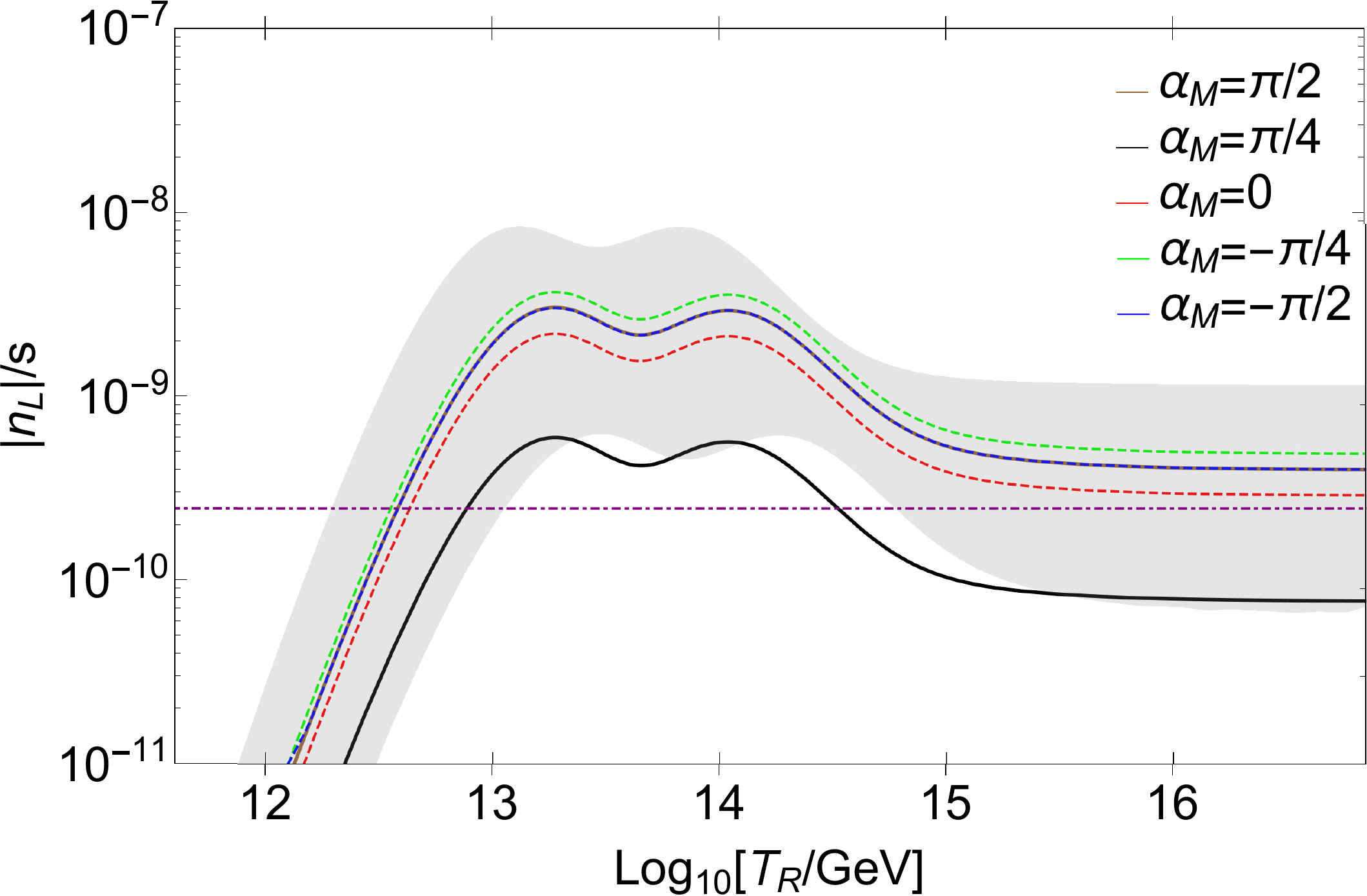}
\end{center}
\caption{
The dependence of lepton asymmetry on the reheating temperature when the inflation main decay channel is Higgs boson with several $\a_M$ with $\d=-\pi/2$. 
The normal and inverted mass hierarchies for neutrinos are shown in left and right panels, respectively. 
$m_\phi/T_R=100$, $B=1$ and $m_{\n \rm lightest}=0 \EV$ are fixed.
The shaded regions denote the uncertainty for $\d=-\pi/2, \a_M=0$ for comparison.
The solid and dashed lines denote the sign of the asymmetry is minus and plus, respectively (the required asymmetry is minus). The lines are obtained by taking $C=C'=1$.
}
\label{fig:Numerical_Result_Higgs}
\end{figure}

\begin{figure}[!t]
\begin{center}  
   \includegraphics[width=75mm]{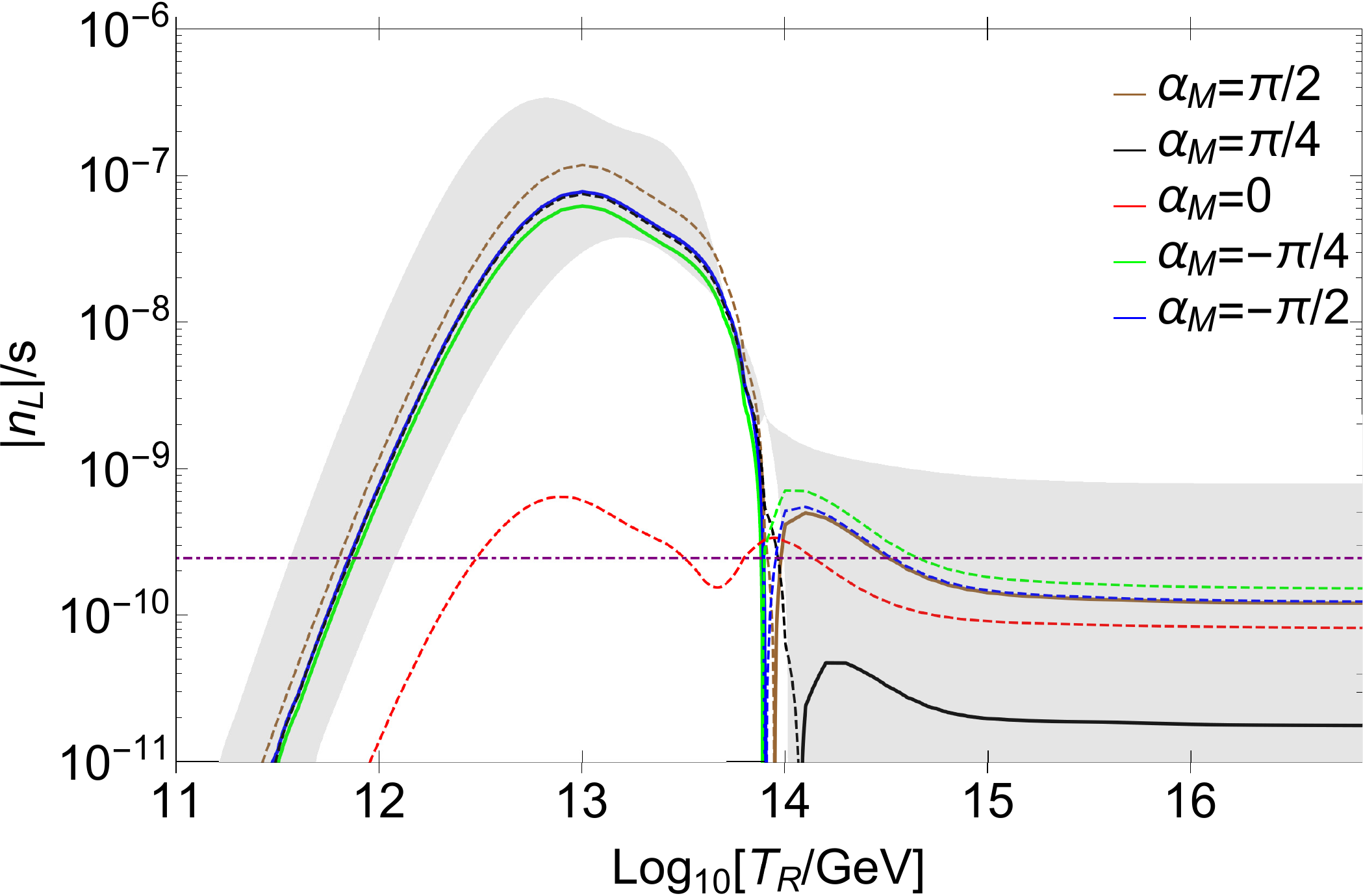}
      \includegraphics[width=75mm]{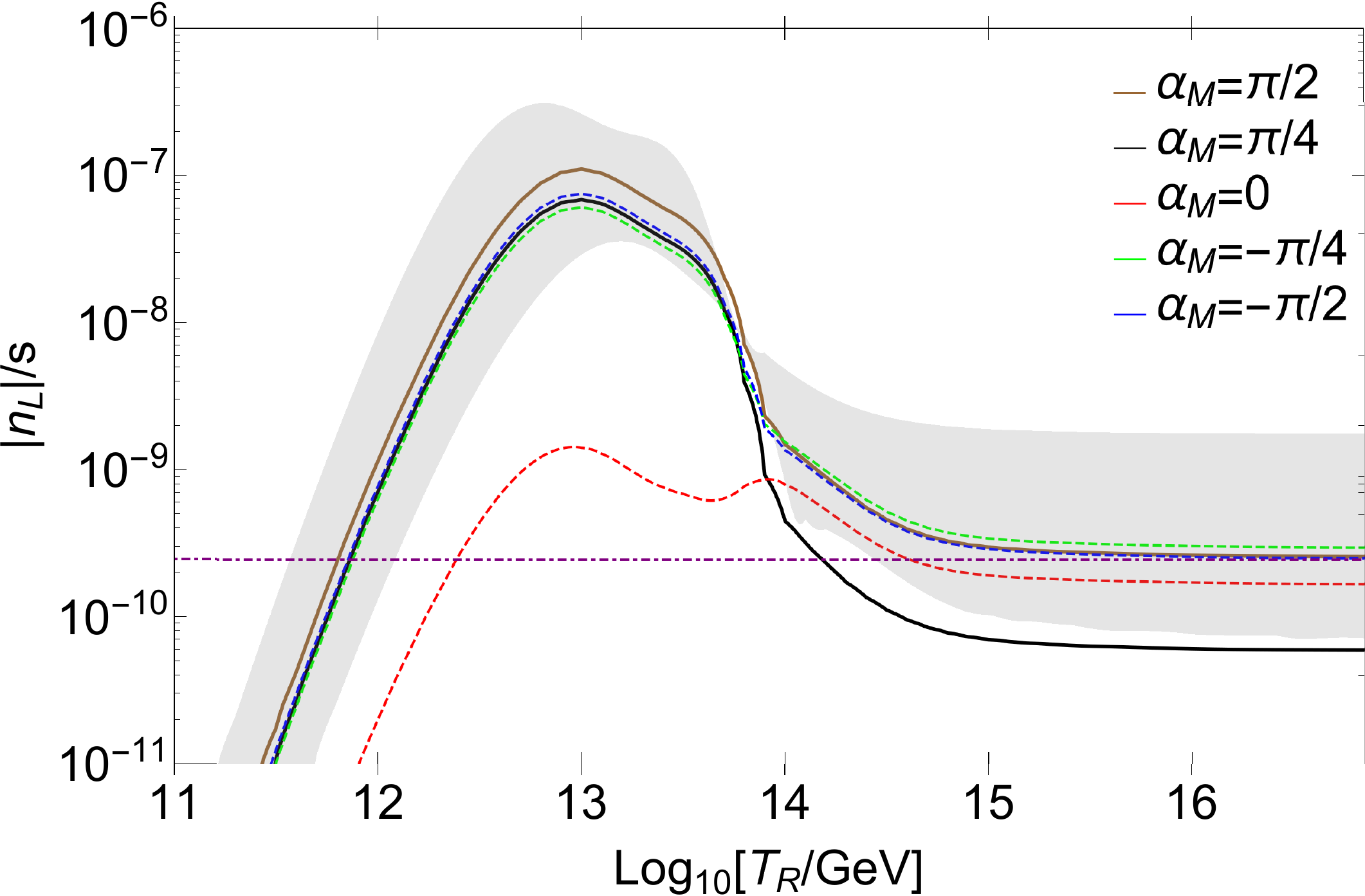}
\end{center}
\caption{
Same as Fig.~\ref{fig:Numerical_Result_Higgs} but the lightest neutrino mass is $m_{\n \rm lightest}=0.07\EV$, and the uncertainty is shown for $\d=-\pi/2, \a_M=\pi/2$ in the gray bands. We take $\a_{M2}=0$.}
\label{fig:Numerical_Result_Higgsdege}
\end{figure}

\begin{figure}[!t]
\begin{center}  
 \includegraphics[width=135mm]{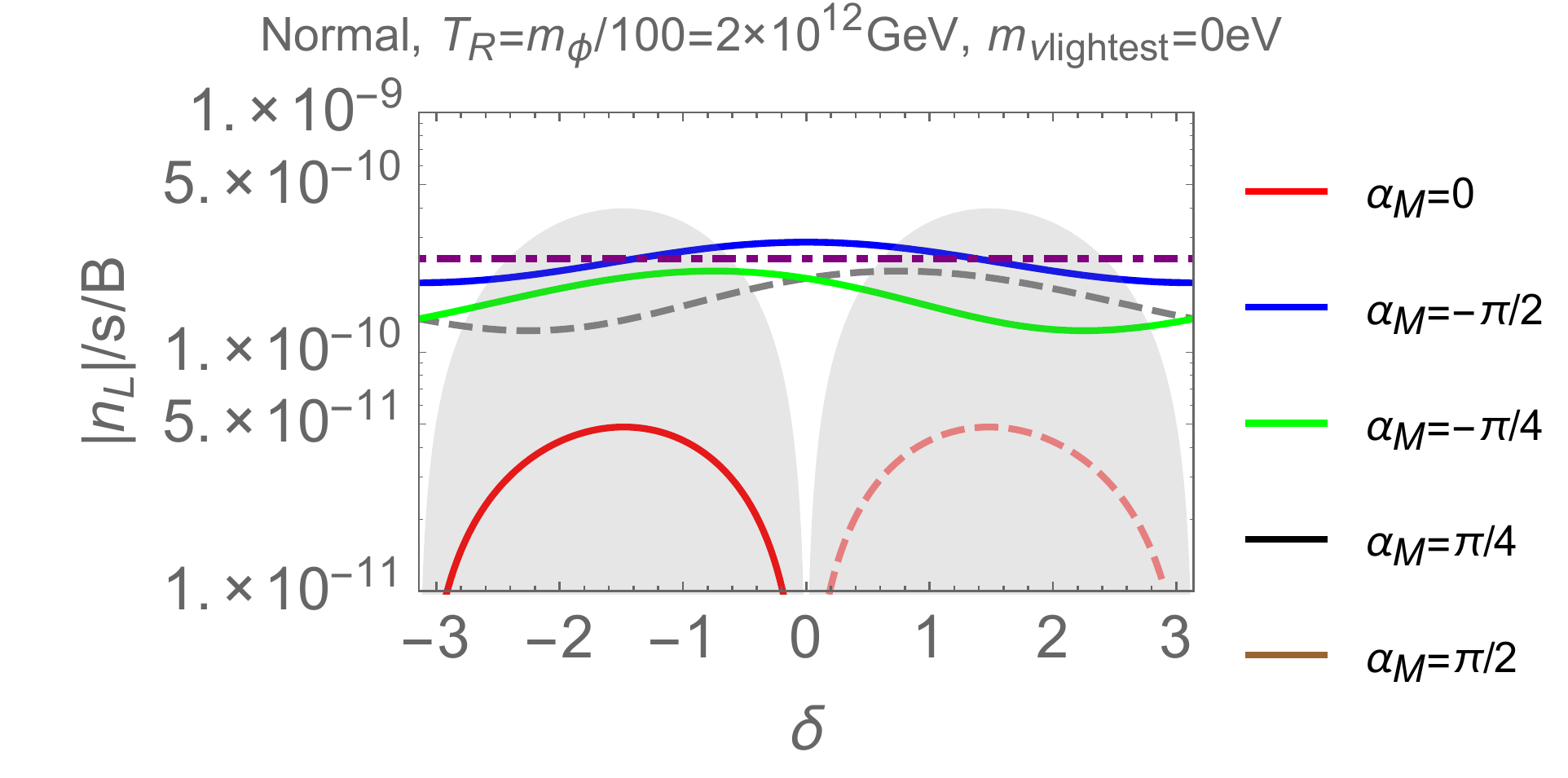}
 \includegraphics[width=140mm]{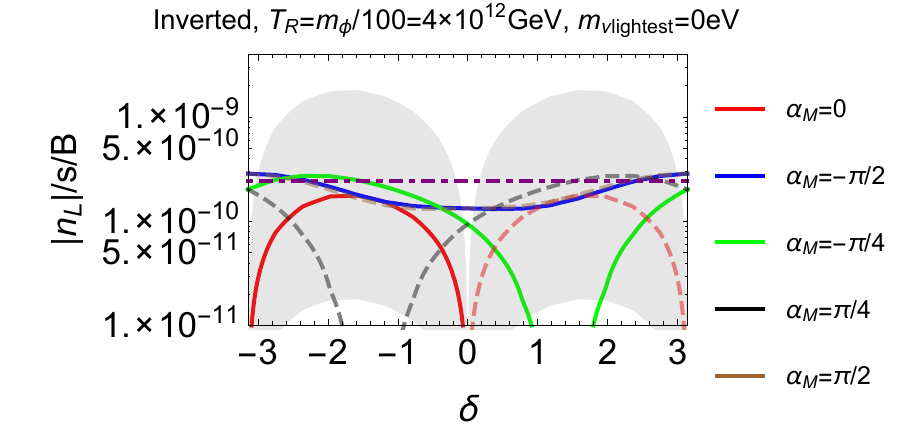}
\end{center}
\caption{
Lepton asymmetry dependence on $\d$ for inflaton decay dominantly to Higgs boson. The uncertainty for $\a_M=0$ case is shown in the gray bands. 
}
\label{fig:deltadeplowTR}
\end{figure}

\begin{figure}[!t]
\begin{center}  
 \includegraphics[width=135mm]{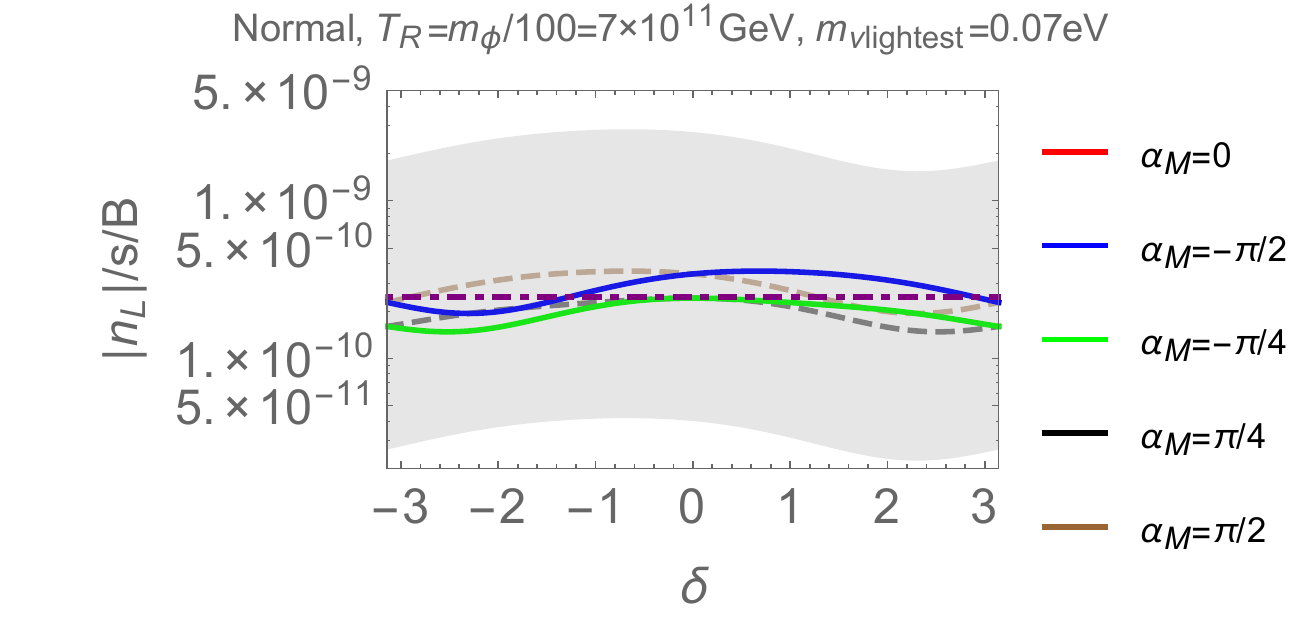}
 \includegraphics[width=140mm]{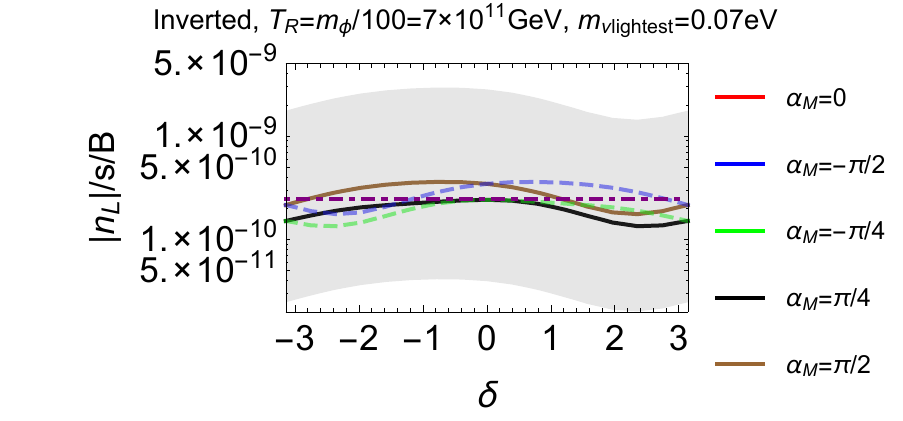}
\end{center}
\caption{
Same as Fig.\ref{fig:deltadeplowTR}, but $m_{\nu {\rm lightest}}=0.07\EV$ is taken with $\alpha_{M2}=0$.  The uncertainty for $\a_M=\pi/2$ case is shown in the gray bands.
}
\label{fig:deltadegedeplowTR}
\end{figure}

\begin{figure}[!t]
\begin{center}  
 \includegraphics[width=115mm]{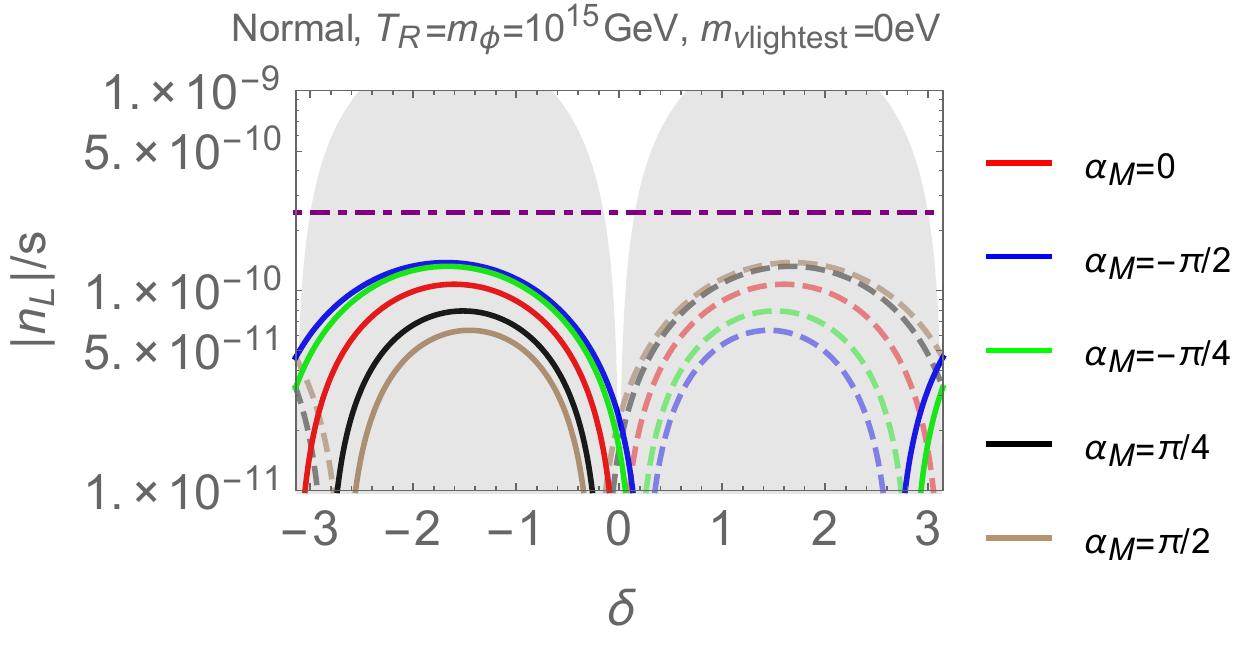}
 \includegraphics[width=115mm]{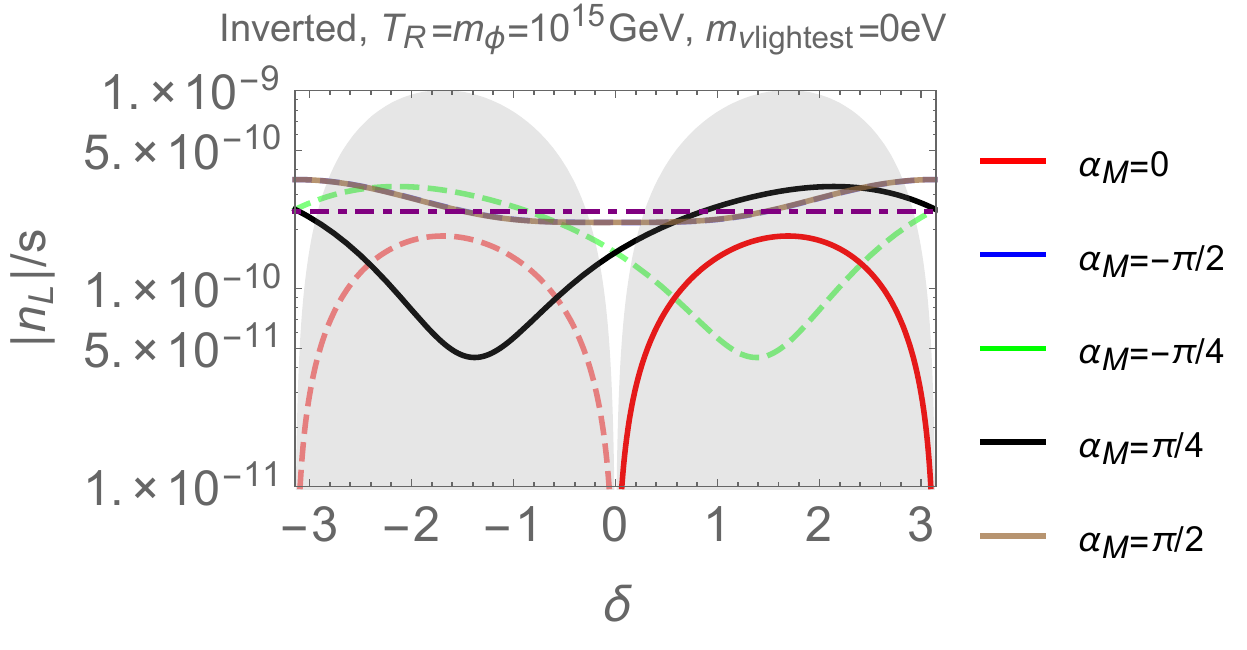}
\end{center}
\caption{
Lepton asymmetry dependence on $\d$ with $T_R=m_\f= 10^{15}\GEV,B=1$, 
for inflaton decay dominantly to Higgs boson. The uncertainty for $\a_M=0$ case is shown in the gray bands. }
\label{fig:deltadep}
\end{figure}

\begin{figure}[!t]
\begin{center}  
 \includegraphics[width=115mm]{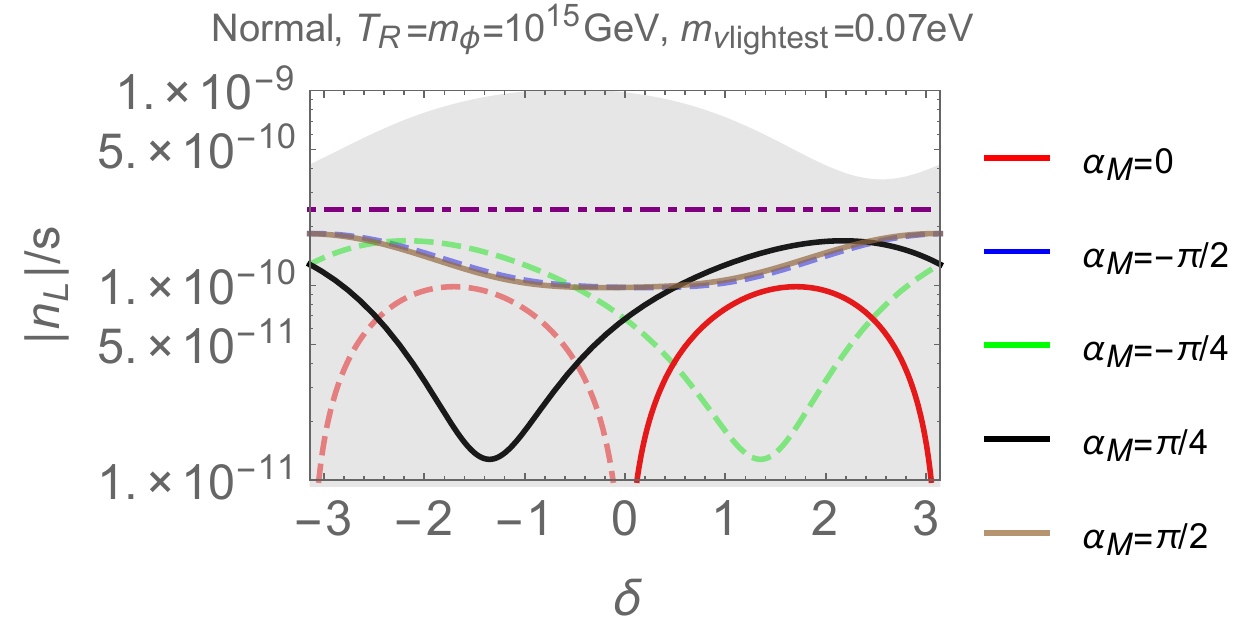}
 \includegraphics[width=115mm]{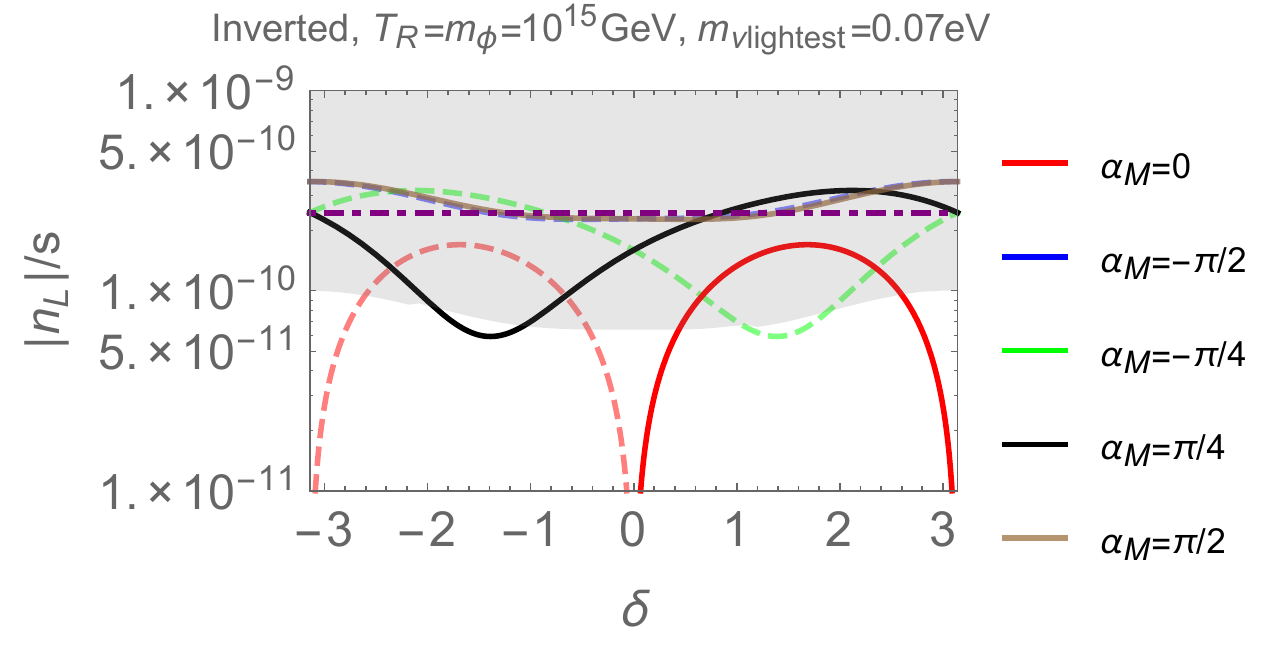}
\end{center}
\caption{
Same as Fig.\ref{fig:deltadep}, but $m_{\nu {\rm lightest}}=0.07\EV$ is taken with $\alpha_{M2}=0$.  
The uncertainty for $\a_M=\pi/2$ case is shown in the gray bands. }
\label{fig:deltadegedep}
\end{figure}

The amounts of the lepton asymmetry in the cases where the inflaton
decays into the Higgs bosons are shown in
Fig.~\ref{fig:Numerical_Result_Higgs} for the hierarchical neutrino mass
cases. The degenerate cases are shown in
Fig.~\ref{fig:Numerical_Result_Higgsdege}. One can see that the large
enough lepton asymmetry is generated for $T_R
\gtrsim10^{11\text{--}12}\GEV$. In this scenario, the sign of the
asymmetry is determined by the parameters in the PMNS matrix. In the
figures, the solid lines represent the good sign (minus), whereas we
draw the dashed lines for the opposite sign.
Different lines correspond to different values of $\alpha_M$ as indicated and $\alpha_{M2} =0$. For other values, the dependence can be found in 
Sec. \ref{sec:analytical_understanding}. We take $C=C'=1$ for those lines.
The bands represent the uncertainties from $C$ and $C'$ for a reference
point $\delta=-\pi/2$ and $\alpha_M=0$ ($\alpha_M=\pi/2$) for Figs. \ref{fig:deltadeplowTR} and \ref{fig:deltadep} (Figs. \ref{fig:deltadegedeplowTR} and \ref{fig:deltadegedep}). We took the same windows of
the uncertainties as before.

\begin{figure}[!t]
\begin{center}  
   \includegraphics[width=75mm]{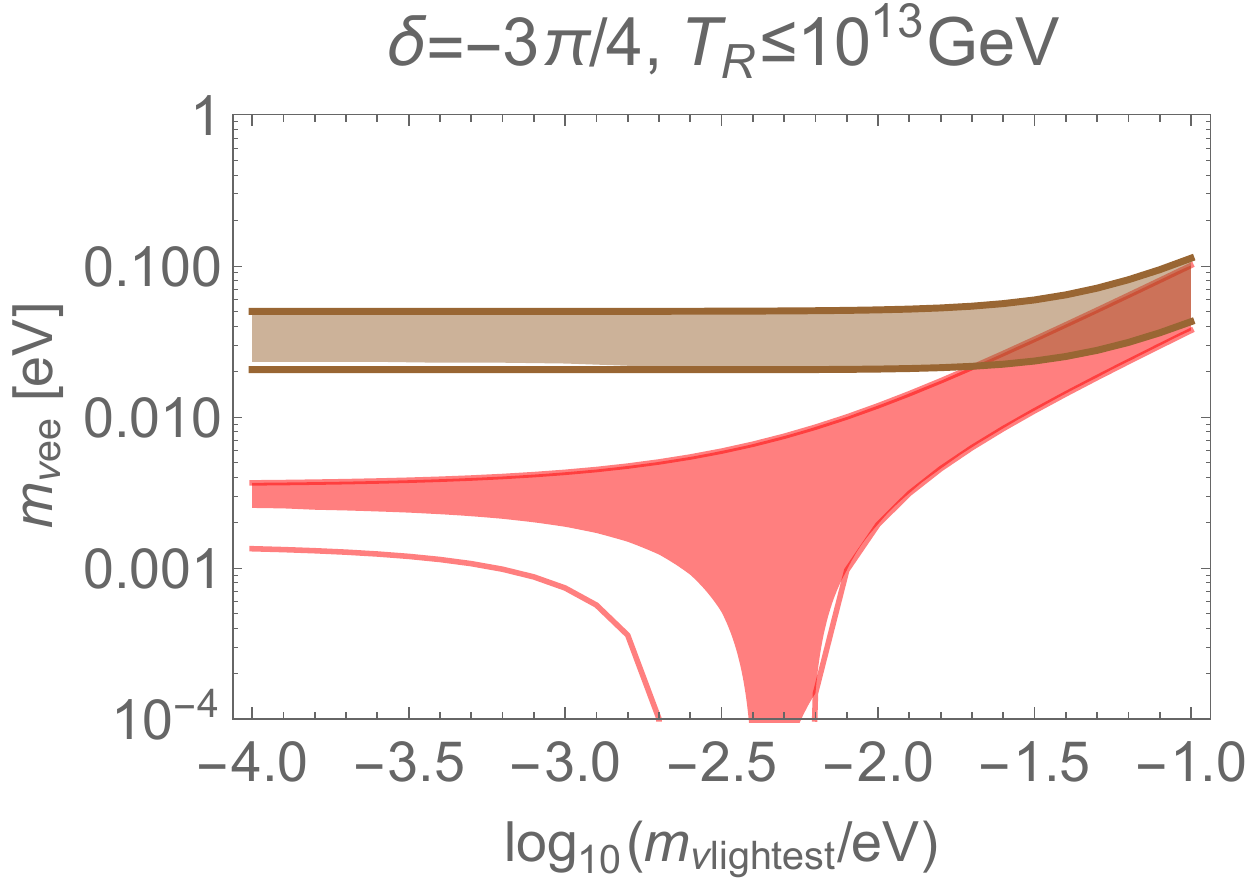}
   \includegraphics[width=77mm]{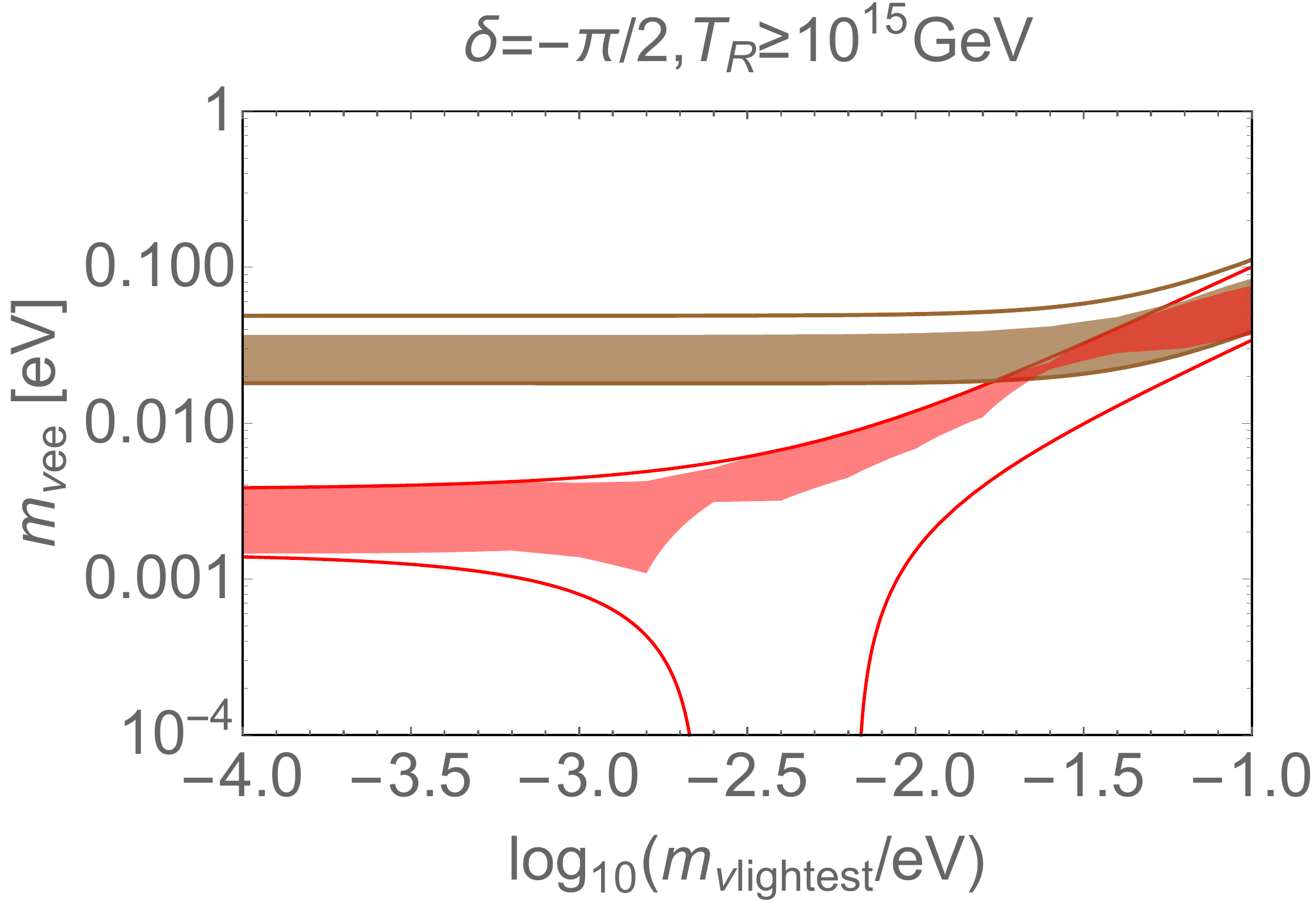}
\end{center}
\caption{
The value of effective neutrino Majorana mass, $m_{\nu ee}$, compatible with our scenario. The inflaton decays into Higgs boson. $T_R\lesssim 10^{13}\GEV, \delta=-3\pi/4$ and $T_R\gtrsim 10^{15}\GEV, \delta=-\pi/2$ are assumed for the left panel and right panel, respectively. The region between upper and lower black (brown) lines is the general possibility for normal (inverted) hierarchy while the shaded regions are our prediction. 
}   
\label{fig:double_beta}
\end{figure}

As we will discuss in Sec. \ref{sec:analytical_understanding}, the mechanisms of the leptogenesis are
qualitatively different for $T_R \lesssim 10^{13}$~GeV and $T_R \gtrsim
10^{14}$~GeV. The dependences on the phases in the PMNS matrices are
shown in Figs.~\ref{fig:deltadeplowTR} (hierarchical) and
\ref{fig:deltadegedeplowTR} (degenerate) for $T_R \sim 10^{12}$~GeV, and
those for $T_R \sim 10^{15 \text{--}16}$~GeV are shown in Figs.~\ref{fig:deltadep} (hierarchical)
and \ref{fig:deltadegedep} (degenerate). 
Since these phases are the parameters in the low energy Lagrangian, one
can check if the predicted sign or amount is consistent with $n_B$ once the
phases are measured in neutrino experiments.
For example, the sign and amount of the measured $n_B$ constrain the allowed region of the
effective neutrino mass $m_{\nu ee}$ for the neutrino-less double beta
decay. An example is shown in the left panel of Fig.~\ref{fig:double_beta}, where we
require that correct sign of the baryon asymmetry is generated with
$T_R\lesssim 10^{13}\GEV$ with a fixed value of $\delta=-3\pi/4$. 
The
requirement reduces the allowed region to 
the shaded one between solid lines. 
The case for $T_R \gtrsim 10^{15}$~GeV and $\d=-\pi/2$ is shown in the right panel. As will be discussed in Sec. \ref{sec:analytical_understanding}, the baryon asymmetry in this case has little dependence on the inflation models, and thus it is more predictive. We require the asymmetry to be within the $1/2-2$ of the measured one, by taking $C=C'=1$.

\section{Underlying mechanisms for leptogenesis}\label{sec:analytical_understanding}

In this section, we discuss how the lepton asymmetries are generated in
each domain of the reheating temperatures and decay modes. We discuss
the following situations separately:
\begin{itemize}
\item inflatons decay into leptons directly and low $T_R$,
\item inflatons decay into Higgs bosons and low $T_R$,
\item inflatons decay into leptons directly and high $T_R$, and
\item inflatons decay into Higgs bosons and high $T_R$,
\end{itemize}
where the separation of high and low $T_R$ is around $10^{13-14}$~GeV as
will be explained later. The mechanisms are qualitatively different in
those four cases.
In Sec.~\ref{Sec:symmetry}, we clarify the necessary conditions for
leptogenesis and how the lepton asymmetry depends on parameters in the
Lagrangian from the symmetry perspective. In Secs. \ref{Sec:case1}, \ref{chap:case2}, \ref{chap:case3}, and \ref{chap:case4}, we discuss each scenario and provide
qualitative and quantitative understandings of the numerical results. 

\subsection{Symmetry argument for CP violation}\label{Sec:symmetry}

In order for enough lepton asymmetry to be generated, CP symmetry must
be broken physically, i.e., the CP phase should not be rotated away in
the interactions that are relevant in the process.  Analogous to the
case of the CP violation in the $K$-meson system, this discussion allows one
to find the most relevant parameters for the asymmetry.

We have possible sources for CP violation: Yukawa
interaction, $llHH$ interaction, and the initial condition. The initial
condition is regarded as the density matrix of inflaton decay product,
and the medium around the reheating temperature.

First, let us consider $T_R \lesssim 10^{14}\GEV$ where the medium is
almost flavor blind since the gauge interactions are more important than
the lepton Yukawa and $llHH$ interactions.
Suppose the limit of two vanishing neutrino masses. The Yukawa
interaction has $\U(1)^3$ symmetry while the $llHH$ interaction has
$\U(2)$ symmetry.  In this limit, one can see that all the CP phases of
the PMNS matrix can be rotated away.  Thus, it implies that either 
\begin{itemize}
\item a CP-odd parameter in the inflaton decay product (in the rotated away basis) or
\item the perturbation of the mass of a lighter neutrino
\end{itemize}
 is needed to
generate the lepton asymmetry.  The lower temperature region of
Figs.~\ref{fig:Numerical_Result} and \ref{fig:Numerical_Result3}
correspond to the former case and the lower temperature region of
Fig.~\ref{fig:Numerical_Result_Higgs} corresponds to the latter case.
In the latter case, $y_\t^2 m_{\n 2} m^*_{\n 3}$ ($ y_\t^2 m_{\n 1}
m^*_{\n 2}$) should appear in the lepton asymmetry for normal (inverted)
mass ordering at the leading order, which comes from the last two terms
of \Eq{deltaGamma}.  Thus the asymmetry is suppressed if $m_{\n 2}$ ($m_{\n 1}$) is
small. Also, we can understand that the Majorana phase is important in this case.

For $T\gg 10^{14}\GEV$, the gauge interactions decouple, and the medium
is not necessarily blind under lepton flavor. The strong $llHH$
interactions at high temperatures quickly bring the initial density matrix in the
diagonal form in the mass basis. 
In the limit of vanishing $y_\mu, y_e$ and the lighter two neutrino
masses while keeping the initial density matrices fixed, the CP phases
of the PMNS matrix can be rotated away by the rephasing of $l_i$ by
$\U(1)^3$ in the mass basis together with the $\U(2)$ rotation in the
flavor basis without changing the initial density matrices.
This implies the final asymmetry should be proportional to either 
\begin{itemize}
\item $y_\tau^2 y_\mu^2$, or
\item $ y_\tau^2
m_{\n 2} m^*_{\n 3}$ ($y_\tau^2 m_{\n 1}m^*_{\n 2}$) for normal
(inverted) mass ordering.
\end{itemize}
These two effects both contribute in the region of high reheating temperatures of Figs.~\ref{fig:Numerical_Result} and~\ref{fig:Numerical_Result_Higgs}.
Possible CP phases in the inflaton decay sector do not contribute since the initial condition is set by the strong $llHH$ interactions.

When the neutrino masses are degenerate, the $llHH$ interaction preserves an $\SO(3)$ symmetry. 
This plays a relevant role as will be seen in Secs. \ref{chap:case3} and \ref{chap:case4}.

 In the following, we discuss two kinds of scenarios depending on whether
the initial condition is in a general matrix (Sec.~\ref{Sec:case1} and latter case of Sec.~\ref{chap:case3}) or
in the diagonal matrix (Secs.~\ref{chap:case2}, \ref{chap:case4} and former case of Sec.~\ref{chap:case3}) in the mass basis.
In both cases, the flavor dependent asymmetries are first generated and
converted into net lepton asymmetry through the lepton-number-violating
$llHH$ interactions. We will see that CP and lepton number violations
are connected through the ``observation'' by the medium.
Since ``observation'' in quantum mechanics is a one-way process, it
provides the departure from the thermal equilibrium, and hence the
Sakharov conditions~\cite{Sakharov:1967dj} are satisfied.

\subsection{Inflatons decay into leptons and $T_R\lesssim10^{13\text{--}14}\GEV$}\label{Sec:case1}

\lac{r1}

For $T_R \lesssim 10^{13-14}$~GeV, the time scale for the thermalization
process,
\begin{align}
 t_{\rm th} & = \Gamma_{\rm th}^{-1}
 \sim 
\left(
{\alpha_2^2 T_R} \sqrt{T_R
 \over m_\phi}
\right)^{-1},
\laq{tth}
\end{align}
is faster than the expansion rate of the Universe, $H (T_R) \simeq 1/t_R$.
Therefore, the high energy component of leptons is continuously produced
by the inflaton decay over the time scale $t_R$, but each lepton loses
the energy very quickly by the time scale $t_{\rm th}$ ($ \ll t_R $).
The scattering processes via gauge interactions do not destroy the
structure of the density matrices. 
After losing their energies, the pair annihilation and pair creation
processes become important. The mean free time of the low energy lepton
is
\begin{align}
 t_{\rm pair}& = \Gamma_{\rm pair}^{-1} \sim 
\left(
{\alpha_2^2 T_R}
\right)^{-1},
\laq{tpair}
\end{align}
which is even shorter than $t_{\rm th}$. Therefore, almost
instantaneously after the inflaton decay, the combination of the density
matrix $\rho_T + \bar \rho_T$ flows to $\rho_T + \bar \rho_T \propto {\bf 1}$ since $ (\d\r_T+\d\ol{\r}_T) \sim e^{- t/t_{\rm pair} }{(\d\r_T +\d\ol{\r}_T)}$ from \Eqs{kinT} and \eq{GammaT}. By
Eq.~\eqref{eq:Deltaeq}, the oscillation is cut-off by the time scale of
$t_{\rm th} + t_{\rm pair} \sim t_{\rm th}$. 

We follow the density matrices of the $\delta \rho_T$ component. Even
though $t_{\rm th} > t_{\rm pair}$ so that the decoherence is faster for
low energy leptons, the oscillation of $\delta \rho_T$ is more important
than that of $\rho_{\bf k}$ since 
$\d \Omega$ is larger for low
energy leptons as in Eq.~\eqref{eq:thermalmass}.
Since the time scale of the Hubble expansion, $t_R$, is longer than
$t_{\rm pair}$ or $t_{\rm th}$, one can ignore the redshift of the
momentum in the following discussion.
The density matrices in the mass basis of neutrinos, $\alpha = 1, 2, 3$,
evolve as
\begin{align}
(\delta \rho_T^{\rm mass})_{\alpha \beta}
& =U_{\rm PMNS}^\dagger (\delta \rho^{\rm flavor}_T)_{ij} U_{\rm PMNS}
\nonumber \\
& =
{\cal N} \, U_{\rm PMNS}^\dagger 
e^{ -i \d \Omega (  |{\bf p}| \sim T  ) t } 
(V V^*) 
e^{ i \d \Omega (  |{\bf p}| \sim T  ) t } 
U_{\rm PMNS},
\label{eq:oscillation}
\end{align}
and
\begin{align}
(\delta \bar \rho_T^{\rm mass})_{\alpha \beta}
& =U_{\rm PMNS}^\dagger (\delta \bar \rho^{\rm flavor}_T)_{ij} U_{\rm PMNS}
\nonumber \\
& =
{\cal N} \, U_{\rm PMNS}^\dagger 
e^{ i \d \Omega (  |{\bf p}| \sim T  ) t } 
(V V^*) 
e^{ - i \d \Omega (  |{\bf p}| \sim T  ) t } 
U_{\rm PMNS}.
\label{eq:oscillationbar}
\end{align}
The thermal corrections $\d \Omega$ in Eq.~\eqref{eq:thermalmass} are
dominated by the ones from the Yukawa interactions in the temperature
range $T_R\lesssim10^{13\text{--}14}\GEV$. 
In this case, the differences in the diagonal components appear if there
is a phase in $U_{\rm PMNS}$ and/or $V$.  By ignoring the electron
Yukawa interaction, one finds
\begin{align}
\tilde \Delta^{\rm mass}_{\alpha\beta}
& :=
\left( \delta \rho^{\rm mass}_T
- \delta \bar \rho^{\rm mass}_T \right)_{\alpha \beta} 
\nonumber \\
&
=
2 i {\cal N}  \bigg[
	\{
	(U_{\rm PMNS})_{e\alpha}^* (U_{\rm PMNS})_{\tau\beta} V_e V_\tau^* 
	-
	(U_{\rm PMNS})_{\tau\alpha}^* (U_{\rm PMNS})_{e\beta} V_e^* V_\tau 
	\}
	\sin \d \Omega_{\tau\tau} t
\nonumber\\&
\phantom{2 i {\cal N}  \bigg[}
	+
	\{
	(U_{\rm PMNS})_{\mu\alpha}^* (U_{\rm PMNS})_{\tau\beta}V_\mu V_\tau^* 
	-
	(U_{\rm PMNS})_{\tau\alpha}^* (U_{\rm PMNS})_{\mu\beta}V_\mu^* V_\tau
	\}
	\sin (\d \Omega_{\tau\tau} - \d \Omega_{\mu\mu} ) t 
\nonumber\\&
\phantom{2  i {\cal N}  \bigg[}
	+
	\{
	(U_{\rm PMNS})_{e\alpha}^* (U_{\rm PMNS})_{\mu\beta}V_e V_\mu^*
	-
	(U_{\rm PMNS})_{\mu\alpha}^* (U_{\rm PMNS})_{e\beta}V_e^* V_\mu
	\}
	\sin \d \Omega_{\mu\mu} t
\bigg].
\label{eq:delta}
\end{align}
For $\alpha=\beta=3$ and $\d \Omega_{\mu\mu}=0$, we obtain
\begin{align}
\tilde \Delta^{\rm mass}_{33}
&
= 2 {\cal N}
\left(
 \cos \theta_{23} \sin 2 \theta_{13} {\rm Im} \left[ e^{- i \delta}
 V_\tau V_e^* \right]
+ \cos^2 \theta_{13} \sin 2 \theta_{23} {\rm Im} \left[ V_\tau V_\mu^*
 \right] 
\right)
\sin \d \Omega_{\tau\tau} t
\nonumber \\
&
= 2 {\cal N}
\left(
0.2 \cdot {\rm Im} \left[ e^{- i \delta}
 V_\tau V_e^* \right]
+ 1.0 \cdot {\rm Im} \left[ V_\tau V_\mu^*
 \right] 
\right) \sin \d \Omega_{\tau\tau} t
\nonumber \\
&
=: 
{\cal N}\, \xi_{CP} \sin \d \Omega_{\tau\tau} t
.
\label{eq:asym}
\end{align}
At this stage, the asymmetry $\tilde \Delta_{33}^{\rm mass}$ is not physical
since it depends on the basis. The trace indeed vanishes. Nonetheless,
in the mass basis, the lepton asymmetry is stored in each neutrino-mass
eigenstate although the net asymmetry is not created.

The finite amount of asymmetry is obtained when we include the effects
of the $llHH$ interaction term. Due to the scattering by this term, the
``observation'' of the neutrino mass basis happens.
The 2 to 2 scatterings by this interaction term reduce or increase the
lepton number by two.
The effects of the $llHH$ interaction can be seen by Eq.~\eqref{eq:trace}. 
The neutrino mass differences imply that the right-hand side is
non-vanishing even if the trace of $\tilde \Delta^{\rm mass}$ vanishes.

The time scale that is important for this $\Delta L = 2$ process is
either $t_R$
or
\begin{align}
\laq{tyuka}
 t_{\rm Yukawa} =
\left(
{9 y_t^2 T_R \over 64\pi^3 } y_\tau^2\right)^{-1}.
\end{align}
The former is the time scale where the temperature is kept $\O(T_R)$,
there the dimension five operators are the most effective, and the
latter is the one for the scattering with the top or bottom quarks
through the tau ($y_\tau$) and the top ($y_t$) Yukawa interactions.
For $T_R \gtrsim 10^{11}$~GeV, $t_{\rm Yukawa} \gtrsim
t_R$.
Beyond $t_{\rm Yukawa}$, the density matrices get diagonal in the flavor
basis by the second term in Eq.~\eqref{eq:GammaT}, which means $\tilde
\Delta$ gets vanishing up to the asymmetry already created. 
Therefore, the creation of the net lepton asymmetry happens with the
efficiency of 
\begin{align}
 \xi_{ll HH}& \sim {21 \zeta (3) T_R^3 \Delta m_{23}^2 \over 16 \pi^3 \langle H
 \rangle^4} \times {\rm min} \left[
t_R, t_{\rm Yukawa}
\right],
\label{eq:efficiency}
\end{align}
for the normal mass ordering. Here, $\D m_{\a\b}^2:=m_{\nu \a}^2-m_{\nu \b}^2$ is the neutrino mass square difference. For inverted, $\Delta m_{23}^2$ is
replaced by $\Delta m_{12}^2$.
 Notice that only the difference of the wash-out effects, and thus the neutrino masses, contributes to the net asymmetry (see \Eq{trace}). The main contribution is, therefore, from the mass difference for the atmospheric neutrino oscillations, $\Delta m^2_{\rm atm} \sim (0.05 \EV)^2$.

The amount of the asymmetry is now estimated as
\begin{align}
 n_L \over s&={{\rm Tr}(\tilde \Delta)}
\sim {T_R \over m_\phi} \cdot B \cdot \xi_{CP} \sin \d \Omega_{\tau\tau} t_{\rm pair} \cdot \xi_{llHH}.
\end{align} 
Putting altogether, we find 
\begin{align}
 {n_L \over s}& \sim 
-2 \times  10^{-6} \cdot \xi_{CP} \cdot B \cdot
\left(
{T_R \over 10^{11}~{\rm GeV}}
\right)^{2}
\left(
{m_\phi \over 10^{13}~{\rm GeV}}
\right)^{-1},
& (T_R \gtrsim 10^{11}~{\rm GeV}),
\end{align}
and
\begin{align}
 {n_L \over s}& \sim 
-2 \times  10^{-6}  \cdot \xi_{CP} \cdot B \cdot
\left(
{T_R \over 10^{11}~{\rm GeV}}
\right)^{3}
\left(
{m_\phi \over 10^{13}~{\rm GeV}}
\right)^{-1},
& (T_R \lesssim 10^{11}~{\rm GeV}),
\end{align}
for normal mass hierarchy.
These well fit the numerical results in Fig.~\ref{fig:Numerical_Result}. 
For other neutrino mass hierarchies, the results are numerically similar as we can see in Figs.~\ref{fig:Numerical_Result} and \ref{fig:Numerical_Result3}.


\subsection{Inflatons decay into Higgs bosons and $T_R\lesssim 10^{13-14}\GEV$}
\lac{case2}
Even if the initial lepton density matrices are diagonal in the flavor basis or in the mass basis, the off-diagonal components of $\Delta$ are generated through the flavor oscillations (c.f. \Eq{delta}). The off-diagonal elements can become physical later due to multiple scatterings. Depending on the basis, the off-diagonal components are actually parts of diagonal components. One interaction tends to eliminate the off-diagonal components in one basis, which may lead to the transferring of the off-diagonal components in the basis into diagonal ones in another basis. Therefore, enough times of scatterings to pick up CP violation can make the off-diagonal component of $\Delta$ made by oscillation physical.  
This transfer of the matrix elements by multiple scattering or observation generally takes place.
The off-diagonal elements in mass basis can be generated through the Hamiltonian with the Yukawa interaction
and could be identified as diagonal components in the flavor basis.
If the $llHH$ interaction is too weak to dump all of the off-diagonal elements, some of the off-diagonal elements in the mass basis would later be observed by the Yukawa interaction as the diagonal elements.
 As a result, if there exists physical CP violation, the flavor-dependent lepton asymmetry is generated through the multiple-observation.

Now consider $T_R\lesssim 10^{13}\GEV$. After the inflaton decay, the off-diagonal components, $\tl{\D}^{\rm mass}_{\a\neq \b}$ are generated through \Eqs{Deltaeq} and \eq{Higgsdecay}.
The flowchart for the dominant multiple-observation process with $T_R\lesssim 10^{13}\GEV$ is as follows: 
\beq
\tilde{\D}_{\a \neq \b}^{\rm mass} \textarrow{\rm last ~ term ~ of ~ \Eq{deltaGamma}}  \textarrow{\rm \text{$\G_{\rm Yukawa}$} }
 \D_{\a = \b}^{\rm mass}  \textarrow{ \text{$ \G_{llHH}$}}  {n_L \o s }.
\eeq
Notice that the last term of \Eq{deltaGamma} should enter, as one can see from the
symmetry discussion for physical CP violation.
It is important that all of the time scales of the interaction are slower
than the expansion.  As a result, each observation is not enough to
remove all the quantum coherence, and the CP violation becomes physical
when enough times of the scattering takes place.  

Let us see the above mechanism by explicit calculation. The kinetic equation is given by
\begin{align}
&i {d \over dt} \tilde{\Delta}^{\rm mass} \simeq 
\left[
\Omega^{\rm mass}, \rho^{\rm mass} + {\bar \rho}^{\rm mass}
\right]
+P(t)\cdot \tilde{\Delta}^{\rm mass},
\end{align}
where we have defined
\begin{align}
\laq{tyuka}
&\Gamma^{\rm Yukawa} :=
{9 y_t^2 T_R \over 64\pi^3 } U^{\*}_{\rm PMNS}\diag{(0,y_\m^2,
 y_\tau^2)} U_{\rm PMNS},
\nonumber\\
&\G_{{llHH}}:=\diag{\(\G_{{llHH},1}, \G_{{llHH},2}, \G_{{llHH},3}\)},
\quad\quad\quad\quad 
\G_{{llHH},\a}:={21\zeta{\(3\)}\o 32 \pi^3} {m_{\n,\a}^2 T^3 \o \vev{H}^4},
\nonumber\\
&P(t)\cdot \tilde{\Delta}^{\rm mass}:=
- {i\over2} \left\{ \G_{\rm Yukawa}+\G_{llHH}, \tilde{\Delta}^{\rm mass}  \right\}
- i{21\zeta{(3)}\o 32\pi^3}\(\k^{\rm mass}\)^*
 \cdot \(\tilde{\Delta}^{\rm mass}\)^t \cdot \k^{\rm mass} \, {T^3},
\end{align}
and have neglected several terms which are not important for the discussion below.
Since we start from the initial condition $\tilde{\Delta}^{\rm mass}=0$ at $t=t_{\rm ini}$, the right-hand side vanishes except for the first term at the early stage. 
The nonzero value of $\tilde{\Delta}^{\rm mass}$ is generated by the first term. 
When the pair production/annihilation by the gauge interaction becomes effective, $\rho^{\rm mass} + {\bar \rho}^{\rm mass}$ gets close to the one proportional to the unit matrix and the first term vanishes. After that, the second term becomes effective. 
From this observation, the equation we should solve is written as 
\begin{align}\label{eq:First_equation}
i {d \over dt} \tilde{\Delta}^{\rm mass}& \simeq 
\left[
\Omega^{\rm mass}, \rho^{\rm mass} + {\bar \rho}^{\rm mass}
\right]+P(t)\cdot \tilde{\Delta}^{\rm mass},
&& (t_{\rm ini}\lesssim t \lesssim t_{\rm cut}),
\\ \label{Eq:Pt_equation}
i {d \over dt} \tilde{\Delta}^{\rm mass}& \simeq 
P(t)\cdot \tilde{\Delta}^{\rm mass}
,
&& (t_{\rm cut}\lesssim t \lesssim t_{\rm end}),
\end{align}
where $t_{\rm cut}$ is the time the oscillation stops, and given by $t_{\rm cut}=t_{\rm ini}+t_{\rm pair}$ for this case.
Eq.~\eqref{eq:First_equation} is easily solved by neglecting the second term, and one gets
\begin{align}
\tilde{\Delta}^{(0)}\simeq -i
\left[
\Omega^{\rm mass}, \rho^{\rm mass} + {\bar \rho}^{\rm mass}
\right]
t_{\rm pair},
\end{align}
where 
\beq
\tilde{\Delta}^{(0)}:=\tilde{\Delta}^{\rm mass}|_{t=t_{\rm cut}}.
\eeq
Note that we can regard $\Omega^{\rm mass}$ and $\rho^{\rm mass} + {\bar \rho}^{\rm mass}$ as constants for $t_{\rm ini}\lesssim t \lesssim t_{\rm ini}+t_{\rm pair}$, and that only the off-diagonal components are generated here.
Up to this stage, no CP violation was necessary. As we discussed before,
the off-diagonal component of $\tl{\D}^{\rm mass}$ is not a CP-odd quantity.
The CP phase in the PMNS matrix can bring this off-diagonal component
into the diagonal entries through the Yukawa and $llHH$ interactions. 
The symmetry argument tells us that the CP phase can be physical when $ y_\tau^2 m_{\n2} m_{\n 3}^*$ appears as a perturbation. 

The solution of the Eq.~\eqref{Eq:Pt_equation} is
\begin{align}\label{Eq:solution}
\tilde{\Delta}^{\rm mass}(t_{\rm end})
=
\mathcal{T}\( e^{-i\int^{t_{\rm end}}_{t_{\rm ini}+t_{\rm pair}} d t'P(t')} \) \tilde{\Delta}^{(0)},
\end{align}
where $\mathcal{T}$ is the time ordered product. The lepton asymmetry can be obtained by taking the trace of the solution.
 In the case of the inflaton decay into Higgs bosons and low $T_R$, 
from Eq.~\eqref{Eq:solution}, the lepton asymmetry is
\begin{align}
\tr\(\tl{\D}\)\simeq
&-{21\zeta(3)\over16\pi^3}
\int^{t_{\rm end}}_{t_{\rm ini}+t_{\rm pair}} dt_1
\int^{t_1}_{t_{\rm ini}+t_{\rm pair}} dt_2
\int^{t_2}_{t_{\rm ini}+t_{\rm pair}} dt_3
\nonumber\\
&\times
\tr\left[
\G_{llHH}(t_1)
\Re\left\{
\G_{\rm Yukawa}(t_2)\(\k^{\rm mass}\)^*\(\tilde{\Delta}^{(0)}\)^t \k^{\rm mass}
\right\}
\right]\(T(t_3)\)^3.
\laq{inthiggslow}
\end{align}
Here $t_{\rm ini}$ corresponds to $t_{\rm ini}={\rm min}(t_R, t_{\rm Yukawa})$, and $\k^{\rm mass} := (U_{\rm PMNS})^t  \kappa U_{\rm PMNS} $. 
One can check that $\O(P^2) \AND \O(P)$ contributions vanish as indicated from the symmetry argument.
By observing that $\G_{llHH}\propto T^3$ and $\G_{\rm Yukawa}\propto T$, the integration is 
dominated by the earlier time, and then one gets
\begin{align}
\tr\(\tl{\D}\)\sim
-{21\zeta(3)\over16\pi^3}t^3 T^3
\tr\left[
\G_{llHH}
\Re\left\{
\G_{\rm Yukawa}\(\k^{\rm mass}\)^*\(\tilde{\Delta}^{(0)}\)^t \k^{\rm mass}
\right\}
\right]\Bigg|_{t=t_{\rm ini}}.
\end{align}

For normal mass hierarchy, $\tilde{\Delta}^{(0)}$ is given by
\begin{align}
&\tilde{\Delta}^{(0)}_{13}=\(\tilde{\Delta}^{(0)}_{31}\)^*\sim
-i\,{\d \Omega^{\rm mass}_{13}\o \G_{\rm pair}}\, \(\rho^{\rm mass} + {\bar \rho}^{\rm mass}\)_{33}
,
&\tilde{\Delta}^{(0)}_{23}=\(\tilde{\Delta}^{(0)}_{32}\)^*\sim
-i\,{\d \Omega^{\rm mass}_{23}\o \G_{\rm pair}}\,\(\rho^{\rm mass} + {\bar \rho}^{\rm mass}\)_{33},
\end{align}
and other components are almost zero, see \Eq{Higgsdecay}. 
The resultant lepton asymmetry is calculated as
\beq
\label{higlow1}
{n_L\o s}\sim 4 \times 10^{-9}{ B} \cdot \x_{CP} \({T_R /m_\f\o 0.01}\)^{1/2} \({T_R \o 10^{13}\GEV }\)^3,~~~~ (T_R \gtrsim 10^{11}\GEV),
\eeq
and 
\beq
\label{higlow2}
{n_L\o s}\sim 1 \times 10^{-20}{ B}\cdot\x_{CP}\({T_R /m_\f\o 0.01}\)^{1/2} \({T_R \o 10^{10}\GEV }\)^6,~~~~ (T_R \lesssim 10^{11}\GEV),
\eeq
where
\beq
\x_{ CP}\sim \(\sin \a_M + 0.2 \sin(\a_M +\d)\) ~~({\rm normal~ hierarchy}).
\eeq
The calculations for inverted and degenerate cases are straightforward, and the results are given by Eqs.~\eqref{higlow1} and \eqref{higlow2} except for the replacement of $\x_{ CP}$:
\begin{align}
&\x_{CP}\sim 0.01 \sin\alpha_M  \cos{\delta}-0.04 \cos\alpha_M \sin{\delta }-0.05 \sin{\alpha_M} ,~~({\rm inverted~ hierarchy}),
\nonumber\\
&\x_{CP}\sim \({\ol{m}^{\rm pole}_\n \o 0.07\EV}\)^4  \(22 \sin ( \a_M -\a_{M2} )-10 \sin ( \a_{M2} ) +4.4 \sin (\alpha_M - \a_{M2} +\delta)\right. \left.+4.5 \sin ( \a_{M2} -\delta) \), \non\\
&({\rm degenerate~ case~ with ~normal ~ordering}), \nonumber\\
&\x_{CP}\sim -\({\ol{m}^{\rm pole}_\n \o 0.07\EV}\)^4  \(22 \sin ( \a_M -\a_{M2} )-9.7 \sin ( \a_{M2} ) +4.5 \sin (\alpha_M - \a_{M2} +\delta)\right. \left.+4.3 \sin ( \a_{M2} -\delta) \), \non\\
&({\rm degenerate~ case~ with ~inverted ~ordering}). 
\end{align}
Here $\ol{m}^{\rm pole}_\n$ means the average of the measured neutrino mass. Here and hereafter the superscript ``pole" is put in order to distinguish $\ol{m}^{\rm pole}_\n$ from the neutrino mass parameter at the high energy scale, see App.~\ref{App:parameters}. These fit the numerical results well in Figs.~\ref{fig:Numerical_Result_Higgs}, \ref{fig:Numerical_Result_Higgsdege}, \ref{fig:deltadeplowTR}, and \ref{fig:deltadegedeplowTR}.

 Unlike \Eq{efficiency}, the lepton number is not proportional to the mass differences of the neutrinos. Even with the equal neutrino masses, asymmetry is generated through the CP violation in the PMNS matrix as one can see in \Eq{inthiggslow}. The CP-violating $llHH$ interactions together with the lepton Yukawa interactions distribute the lepton number into left and right-handed leptons while net asymmetry vanishing. The asymmetry stored in the left-handed leptons are partially washed out by the first term in \Eq{trace}, and the net asymmetry is generated.

\subsection{Inflatons decay into leptons and high $T_R$}
\lac{case3}

At a high-temperature range, 
the lepton asymmetry is generated through multiple ``observations'' of
leptons in the medium as in the previous subsection. 
Contrary to the previous subsection, the leptons are observed at different temperatures and thus in different 
basis.  As a result, non-observable
off-diagonal components in one basis can later be observed as diagonal
components in another basis. 

\paragraph{Case with hierarchical neutrino masses}
Here we consider the high-temperature regime where $T_R \gtrsim 10^{15}\GEV$ for the normal and inverted hierarchy with one massless neutrino.\footnote{Although \Eq{pert} is satisfied in the effective theory at $T_R\lesssim 10^{16}\GEV$, the UV physics might contribute to the following mechanism. As we will see, these contributions do not change our prediction much, if we assume that the lepton number for the massless neutrino is not violated in the UV physics at the vanishing limit of the Yukawa couplings.} In this region, the numerical calculation shows an interesting feature that the asymmetry gets almost independent of the reheating temperature.

For simplicity, we consider $ \ab{\bf k} \sim T$, in which case $t_{\rm th}
\sim t_{\rm pair}$.  It is useful to define the following density
matrix, \beq \laq{largeT} \(\r^{\rm mass}_T\)_{\a\b}\equiv
\int_{\ab{{\bf p}}\sim T, \ab{{\bf k}}}{{d^3{\bf p}\over (2\pi)^3}
{\r^{\rm mass}_{\a \b}({\bf p},t)\over s}}.  \eeq

At $T_R \gtrsim10^{15}\GEV$, the time scale of $t_{\rm pair},t_{\rm th}$
and $t_{\rm Yukawa}$ are slower than the expansion rate $T_R$, while the $llHH$ interactions are faster than $t_R$, and %
these are the interactions to
bring the momentum distribution to the thermal one. The discussion is,
therefore, qualitatively different from the case with lower reheating
temperatures. The
density matrix is diagonal in the mass basis due to the fast $llHH$
interactions.

Because of the hierarchy, the thermalization is effective only for two
of the neutrino generations. The density matrices for the leptons has
the following form for $T \gtrsim 10^{15}$~GeV:
\beq 
\r_T^{\rm mass} \sim \diag{(\r_1 (T),\r_2 (T),\r_3 (T))}.
\eeq
where $\rho_1 \neq \rho_2 \sim \rho_3$ for the normal hierarchy and
$\rho_1 \sim \rho_2 \neq \rho_3$ for the inverted one.  
The off-diagonal components are highly suppressed due to the decoherence effect via the $llHH$ interaction (see \Eqs{kinK} and \eq{kinT}). 
The density matrices are kept in this form until the $llHH$ interaction
for the second heaviest neutrino gets ineffective at $T_{\rm ini} \sim
10^{15}$~GeV ($10^{13}$~GeV) for the normal (inverted) hierarchy.
Here $T_{\rm ini} $  is the temperature $\G_{llHH,2}=H(T)$ ($\G_{llHH,1}=H(T)$) for normal (inverted) mass hierarchy.

Below the temperature, $T_{\rm ini} $, the off-diagonal components are started to be generated by flavor oscillation (see \Eq{Deltaeq}). The differences among $\rho_1$, $\rho_2$, and $\rho_3$ are important for this to happen.
Since the $llHH$ interaction for the heaviest neutrino is still effective, the oscillation can only generate the $\tl{\Delta}^{\rm mass}_{12}$ ($\tl{\Delta}^{\rm mass}_{13}$) component for the normal (inverted) mass hierarchy. 
The oscillation continues until the time scale that the pair annihilation by the gauge interactions becomes as fast as the expansion rate.  
Even after the gauge interaction rate becomes faster than the expansion rate, the generated off-diagonal element is kept unerased in the medium due to the flavor blindness of the gauge interactions. 
Finally, when the Yukawa interaction becomes effective, $\tilde t_{\rm Yukawa} \sim M_{\rm P}/3T_\tau^2$ with $T_\tau \sim10^{11}$~GeV which is the time ${9 y_t^2 y_\tau^2 \o 64\pi^3}  T_\tau=H(T_\tau)$, the density matrix gets diagonal in the flavor basis. The generation of the asymmetry stops at this time.

For normal mass hierarchy case, there exist the contributions which depend on $y_\tau^2 y_\mu^2$ and $y_\tau^2 m_{\nu2}m_{\nu3}^*$, respectively.
Flowcharts to describe the dominant processes for leptogenesis can be drawn
as 
\beq 
\tl{\D}^{\rm mass}_{12} \textarrow{ \text{$\Gamma_{\rm Yukawa}
$}}\textarrow{ \text{$\G_{\rm Yukawa}$} } \tl{\D}^{\rm mass}_{33}
\textarrow{\text{$\G_{llHH}$} } {n_L \o s },
\eeq
for $y_\tau^2 y_\mu^2$ contribution, and
\beq
\tl{\D}^{\rm mass}_{12} \textarrow{ \text{$\G_{\rm Yukawa}$} }\textarrow{ {\rm  last ~term~ of~ \Eq{deltaGamma} } } \textarrow{ \text{$\G_{\rm Yukawa}$} } \tl{\D}^{\rm mass}_{33} \textarrow{\text{$\G_{llHH}$}  }  {n_L \o s },
\eeq
for $y_\tau^2 m_{\nu2}m_{\nu3}^*$ contribution.

From the above discussion, we can take $t_{\rm ini}\sim M_{\rm P}/3T_{\rm ini}^2 $, $t_{\rm cut}= \tl{t}_{\rm pair}$ and $t_{\rm end}=\tilde{t}_{\rm Yukawa}$, where $\tl{t}_{\rm pair}$ is the time scale at which the gauge interactions are imporant, $\alpha_2^2 T \simeq H(T)$.
The form of the density matrix at $t=t_{\rm ini}$ is
\begin{align}
&\r_T^{\rm mass}=\bar{\r}_T^{\rm mass}=\diag{(\r_1,\r_2,\r_3)},
&&\r_1\neq \r_2 =\r_3,
\end{align}
for the normal mass hierarchy. 
Since the strong $llHH$ interactions bring the densities to the thermal ones except for $\rho_1$, 
\beq \rho_2 = \rho_3\simeq {0.004 \({100\o
g_{*s}(T_R)}\) \delta_{ij}}, \eeq
\beq \rho_1 - \r_2 \sim  {3\o 4}B \ab{V_i \(U^*_{\rm PMNS}\)_{i 1}}^2 .\eeq
Then, we obtain by solving \Eq{First_equation}
\begin{align}
\tilde{\Delta}_{12}^{(0)}
=
\(\tilde{\Delta}_{21}^{(0)}\)^*
=
2i (\r_1-\r_2) {\Omega^{\rm mass}_{12}  \o \G_{\rm pair}},
\end{align}
and vanishing other components at $t=\tl{t}_{\rm pair}$. 
Here $\tl{\D}^{(0)}_{13}=(\tl{\D}^{(0)}_{31})^*,\AND \tl{\D}^{(0)}_{23}=(\tl{\D}^{(0)}_{32})^*$ are negligible because the fast decoherence at the rate ${1\o2} \G_{llHH,3}$. 
$\tl{\D}^{\rm mass}_{12}=(\tl{\D}^{\rm mass}_{21})^*\simeq \tilde{\Delta}_{12}^{(0)}$ does not change much until $t=\tl{t}_{\rm Yukawa}.$
From Eq.~\eqref{Eq:solution}, the lepton asymmetry is
\begin{align}\label{Eq:Dirac_phase_contribution}
\tr\(\tl{\D}\)\bigg|_{y_\tau^2y_\mu^2}\simeq
-
\int^{t_{\rm end}}_{t_{\rm ini}+t_{\rm pair}} & dt_1
\int^{t_1}_{t_{\rm ini}+t_{\rm pair}} dt_2
\int^{t_2}_{t_{\rm ini}+t_{\rm pair}} dt_3 
\nonumber\\
&\times
\,\tr\left[
\G_{llHH}(t_1)
\Re\left\{
\G_{\rm Yukawa}(t_2) \tilde{\Delta}^{(0)} \G_{\rm Yukawa}(t_3)
\right\}+...
\right],
\end{align}
for ${y_\tau^2y_\mu^2}$ contribution, and is  
\begin{align}\label{Eq:Majorana_phase_contribution}
\tr\(\tl{\D}\)\bigg|_{y_\tau^2 m_{\nu2}m_{\nu3}^*}\simeq
&{21\zeta(3)\over16\pi^3}
\int^{t_{\rm end}}_{t_{\rm ini}+t_{\rm pair}} dt_1
\int^{t_1}_{t_{\rm ini}+t_{\rm pair}} dt_2
\int^{t_2}_{t_{\rm ini}+t_{\rm pair}} dt_3
\int^{t_3}_{t_{\rm ini}+t_{\rm pair}} dt_4
\nonumber\\
&\times
\,\tr\left[
\G_{llHH}(t_1)
\Re\left\{
\G_{\rm Yukawa}(t_2) 
\(\k^{\rm mass}\)^*\(\G_{\rm Yukawa}(t_4) \tilde{\Delta}^{(0)} \)^t \k^{\rm mass}
\right\}
+...\right]\(T(t_3)\)^3,
\end{align}
for $y_\tau^2 m_{\nu2}m_{\nu3}^*$ contribution. 
... are the subdominant terms for the normal mass hierarchy from the anti-commutation in $P$.
One can see that Eq.~\eqref{Eq:Dirac_phase_contribution} is dominated by the large $t$ region while all range of $t$ equally contributes to the integral in Eq.~\eqref{Eq:Majorana_phase_contribution}.
As a result, we obtain 
\begin{align}
{n_{L} \o s}\bigg|_{y_\tau^2y_\mu^2} &\sim \left. \,  -\G_{llHH,3} \times \Re{[\G_{31}^{\rm Yukawa}
 \tl{\D}_{12}^{(0)} \G_{23}^{\rm Yukawa}]}  t^3 \right|_{t=\tl{t}_{\rm Yukawa}} \non\\
&\sim 
 - 6\times 10^{-8}\sin\d \times \({T_\tau \over 10^{11}\GEV}\)^2 \times  \(\r_1-\r_2\),
 \laq{HT1}
\end{align}
for $y_\tau^2 y_\mu^2$ contribution, and
\begin{align}
\laq{HT2}
{n_{L}\o s}\bigg|_{y_\tau^2 m_{\nu2}m_{\nu3}^*}  &\sim  \left. \G_{llHH,3}\times {21 \zeta{\(3\)} \o 16\pi^3} {m_{\n 2} m_{\n 3}\o \vev{H}^4} T^3 \times \Re{[\G_{31}^{\rm Yukawa} \tl{\D}_{12}^{(0)} \G_{32}^{\rm Yukawa}]}   t^4 \right|_{t=\tl{t}_{\rm Yukawa}} \non \\ 
& \sim -4\times 10^{-9}\(\sin{\alpha_M }  +0.2 \cos{(\delta+\a_{M}) } \) \(1-0.4 \cos{\d}\) \times \(\r_1-\r_2\).
\end{align}
for $y_\tau^2 m_{\nu2}^*m_{\nu3}$ contribution. 
These formulas well fit the numerical results in the left panel of Fig.~\ref{fig:Numerical_Result}.

Now let us comment on the cases for this mechanism with inverted mass hierarchy.  
The same discussion applies by making exchanges between the indices
$\a,\b=1,2,3$ and $\a,\b=3,1,2 $ in the previous discussion.
However, since two of the $llHH$ interactions are strong,
$\tl{\D}_{31}^{0}=(\tl{\D}_{13}^{0})^*$ is soon destroyed by the $llHH$ interaction, and the
final asymmetry is suppressed. 
On the other hand, we will see soon that if the reheating temperature is slightly smaller than $ 10^{15}\GEV$, an approximate symmetry preserves $\tl{\D}_{12}^0$ and a sufficient amount of the lepton asymmetry can be generated.

One of the essences of this region is the fact, $\r_T+\ol{\r}_T$ is not proportional to the unit matrix and does not commute with $\Omega_T$ at $T>T_{\rm th}$. 
The mechanism here works in general:
e.g. the thermal decoupling of right-handed neutrinos at $T>T_{\rm th}$
would lead to $\r_T + \ol{\r}_T$ not proportional to $\bf{1}$.

\paragraph{Case with degenerate neutrino masses} Let us consider the degenerate case, ${m^2_{\n \a}}\sim {\ol{m}^2_{\n
}} \gg {\ab{\D m^2_{\a\b}}}$, and the time scales of the $llHH$ interaction are faster than $t_{\rm
R}$. Naively, it is expected that the lepton number generated is soon washed out. However, one finds that the imaginary
part of the $\tl{\D}^{\rm mass}$ is not affected during the scattering
via the $llHH$ interaction in \Eqs{GammaT} and \eq{deltaGamma}.  This can be
understood from an approximate $\SO(3)$ lepton flavor symmetry in the
$llHH$ interaction.  The generators are $i \e_{\a\b\g} (a^\*_\a a_\b+b^\*_\a
b_\b)$. Thus the combination of the density matrices,
 \beq 
\Im{[\tl{\D}^{\rm mass}_{\a\b}]}=-i \vev{a_\b^\*a_\a-b^\*_\a b_\b-a_\a^\*a_\b  +b^\*_\b b_\a }/2V, 
 \eeq
 conserves.
 Since the $llHH$ interaction rate is faster than the expansion rate, a non-vanishing $\tl{\D}^{\rm mass}$ quickly flows to the following form
\beq
\tl{\D}^{\rm mass}_{\a\b}\simeq i \Im[{\tl{\D}^{\rm mass}_{\a\b}}].
\eeq
Notice that this symmetry property would be missed in the ordinary Boltzmann equation.

The finite mass differences break the $\SO(3)$ symmetry, and from Eq.~(\ref{Eq:Pt_equation}) the decoherence of the
imaginary part happens at a rate
\beq
\laq{offdiag}
{d \o dt}{\Im{[\tl{\D}}^{\rm mass}_{\a\b} ]}
\sim
- {21\zeta{\(3\)}\o 64\pi^3} {(m_{\n\a}-m_{\n\b})^2\o \vev{H}^4} T^3 \Im{[\tl{\D}}^{\rm mass}_{\a\b} ].
\eeq
When the coefficient in the r.h.s becomes faster than the cosmic
expansion, the imaginary part becomes almost zero.%

Now, for simplicity let us consider the normal mass ordering with
$\ol{m}_\n \sim \O(0.1) \EV$ and $10^{13}\GEV \ll T_R\ll 10^{16}\GEV$ as
an example. 
In this case, $\tl{\D}^{\rm mass}_{ 31}$, $\tl{\D}^{\rm mass}_{ 32}$ and
$\Re{[\tl{\D}^{\rm mass}_{ 31}]}$ quickly go to zero while the
$\SO(2)$ symmetry preserves $\Im{[\tl{\D}^{\rm
mass}_{12}]}$.  
The high energy leptons from the inflaton decays are scattered into medium and go on oscillating at a time scale $\ol{t}_{llHH}=\({\ol{m}_\n^2 \over \vev{H}^4}{21\zeta{\(3\)} T_R^3\o 32\pi^3}\)^{-1}$. The oscillation stops at $\ol{t}_{llHH}$ because the $llHH$ interaction with degenerate neutrino masses also brings $\r_T+\ol{\r}_T$ to be proportional to the unit matrix.
The relevant component from the oscillation is given by 
\beq
\laq{sd2}
\tl{\D}^{\rm mass}_{12}\sim i{T_R \over m_\phi} \cdot B \cdot \x_{ 12}   {\(\d \Omega^{\rm mass}_{22}(T_R)-\d \Omega^{\rm mass}_{11}(T_R)\) (\ol{t}_{llHH})} \sim 2 i \({T_R \over m_\phi}\) \cdot B \cdot  \x_{ 12} \({\D {m^2_{21}} \o \ol{ m}^2_{\n}}\),
\eeq
with
\beq
\xi_{\rm 12}:=    (U^*_{\rm PMNS})_{ i 1}  (U_{\rm PMNS})_{j2}V_iV^*_j.
\eeq
Notice that we have used the oscillation term at $\ab{\bf p}\simeq T_R$ which is dominant as in \Eq{thermalmass}. 
The off-diagonal element is produced with a strong phase but the real part quickly approaches to zero due to the decoherence, which results 
\beq
\tl{{\D}}_{12}^{(0)}=(\tl{{\D}}_{21}^{(0)})^*=i \Im{[\tl{\D}^{\rm mass}_{12}]},
\eeq
and almost vanishing other components in $\tl{\D}^{(0)}.$ $\tl{{\D}}_{12}^{\rm mass}=(\tl{{\D}}_{21}^{\rm mass})^*$ is almost frozen until $t=\tl{t}_{\rm Yukawa}$ due to the symmetry protection.

As noted, although $\tl{\D}^{(0)}$ has vanishing diagonal components in the mass basis, diagonal components in the flavor basis can be non-zero. This implies $\tl{\D}^{(0)}$ is distributed by Yukawa interaction into the left-handed and right-handed leptons from \Eq{trace}: 
\beq
\tr{(\tl{\D}_T)} \rightarrow \tr{(\tl{\D}_T)}-\tr{(\d \tl{\D})}, ~\tr{ (\tl{\D}_R)} \rightarrow  \tr{( \tl{\D}_R)}+\tr{(\d \tl{\D})}
\eeq
where
\beq
\tr{(\d\tl{\D})}=  2{dt}  \Im{[\tl{\D}^{(0)}_{12}]} \Im{[(\G_{\rm Yukawa}(t))_{12}]}
\eeq
for a very short time range $dt$.
However, $-\tr{(\d \tl{\D})}=-\Re[\d \tl{\D}]$ in the left-handed leptons is quickly washed out,  while the one in the right-handed leptons remains:
\beq
\tr{(\tl{\D}_T)} \rightarrow \tr{(\tl{\D}_T)}, ~\tr{ (\tl{\D}_R)} \rightarrow  \tr{( \tl{\D}_R)}+\tr{(\d \tl{\D})}.
\eeq
Therefore the net asymmetry is generated 
and stored in the right-handed leptons. 
The net asymmetry can be obtained from the integration, 
\beq
{n_L \o s}\simeq \int^{t_{\rm end}}_{\ol{t}_{llHH}}{ dt 2\Im{[\tl{\D}^{(0)}_{12}]} \Im{[(\G_{\rm Yukawa}(t))_{12}]}},
\eeq
where $t_{\rm end}=\tl{t}_{llHH}$ is the time $\G_{llHH,1}\simeq \G_{llHH,2}\simeq H(T)$, up to when the net asymmetry is efficiently produced due to the wash-out effect. We obtain
\begin{align}
{n_L \o s} &\sim  \left. 2 \Im{[\tl{\D}^{(0)}_{12}]} \Im{[(\G_{\rm Yukawa})_{12}]} t \right|_{t=\tl{t}_{llHH}} \non \\
&\laq{asydege} \sim - 5\times 10^{-6}\(\sin{\a_M \o2}+0.3\cos{\a_M \o 2} \sin \d \) B  \Re{[\xi_{\rm 12}]}  \({T_R /m_\phi \over 0.01}\) \({\(\D m_{21}^2\)^{\rm pole}\o (0.009\EV)^2}\).
\end{align}
The result does not depend much on $\ol{m}_\n$. 
A same discussion can be applied to the inverted ordering case at the same range of reheating temperature. In particular, the approximate $\SO(2)$ symmetry even works with the lightest neutrino massless.  
The behavior can be found in Fig.~\ref{fig:Numerical_Result3} and the right panel of Fig.~\ref{fig:Numerical_Result}.

\subsection{Inflatons decay into Higgs bosons and $T_R\gtrsim 10^{14}\GEV $}\lac{case4}
When $T_R\gtrsim 10^{14}\GEV$, the gauge interaction decouples and the scattering and thermalization are made by some/all of the $llHH$ interactions. 
One can see the asymmetry approaches to UV insensitive values for all the cases. Two kinds of mechanisms are operating for these UV insensitive values depending on the neutrino mass hierarchies. The dominant asymmetry is not from $\r_{\bf k}$ and $\ol{\r}_{\bf k}$ in the kinetic equation. This is because the dominant oscillation frequency is $\propto T_R$, but it is cutoff by $llHH$ interactions whose time scales are proportional to $T_R^{-3}$. In total, together with the yield of the high energy leptons, $\propto B T_R/m_\phi$, the generated asymmetry is proportional to $T_R^{-1}$, and thus it is suppressed at large $T_R$. Therefore the dominant asymmetry comes from the thermalization process in the medium. 

\paragraph{Normal mass hierarchy}
For the normal mass hierarchy with one massless neutrino, at $T_R\gg10^{15}\GEV$ the asymmetry becomes UV insensitive as in \Sec{case3}.
Since the leptons in the medium are thermalized through the $llHH$ interaction,  the lepton density of the medium has the form
\beq
\r_T^{\rm mass}(T_R)\sim \diag{\(0,0.04 \({11\o g_{*s}(T_R)}\),0.04 \({11\o g_{*s}(T_R)}\)\)}.
\eeq
Notice that at this reheating temperature, only Higgs boson and two of the left-handed leptons are thermalized. 
From \Eq{HT1},
one obtains the dominant asymmetry
\begin{align}
{n_L \o s} \sim & 2\times 10^{-9}  \({11\o g_{*s}(T_R)}\)  \sin\d  \({T_\tau \over 10^{11}\GEV}\)^2 .
\end{align}
The observed asymmetry favors $\d< 0$. 
This formula fits well the results of normal ordering cases in Fig. \ref{fig:Numerical_Result_Higgs} and \ref{fig:deltadep}.
Notice that in the numerical calculation we have conservatively taken $g_{*s}= 100$. More realistic treatment of $g_* \AND g_{*s}$ may give larger asymmetry than the numerical one. 

\paragraph{Inverted mass hierarchy and degenerate masses}
At $T_R\gtrsim 10^{14}\GEV$ for the degenerate cases or the inverted mass hierarchy case, the UV insensitivity also appears. 
The key fact is the departure from the thermal equilibrium of the right-handed leptons with $T\gtrsim 10^{14}\GEV$, where the pair creation rate of $\U(1)_Y$ is smaller than the expansion rate of the Universe. 
Just above $T\simeq 10^{14}\GEV$, there are three (two) generations of left-handed leptons, the Higgs bosons and tops are thermalized for degenerate (inverted mass hierarchy) case due to the $llHH$ and top Yukawa interactions.\footnote{Depending on the uncertainty of the gauge interaction rates, there could also be other particles.}
 The Yukawa interaction, whose rate is much slower than the expansion rate of the Universe, tends to thermalize the right-handed leptons through, for example, $l_\t$-top scattering into right-handed tau lepton and top. 
However, the inverse-process is suppressed due to the absence of thermalized right-handed leptons. In total, the amount of the left-handed leptons are decreased from the thermal equilibrium due to the scattering.
This implies that at $t<t_{\rm ini}+\tl{t}_{Y}$ ($\tl{t}_Y$ is the time at $\a_Y^2T=H(T)$), the deviation from thermal equilibrium, $\d\r^{\rm mass}+\d\ol{\r}^{\rm mass}$, is produced at a rate (See also \Eq{kinT} and App. \ref{App:right-handed_leptons}):
\begin{align}
\laq{source}
{d  \o dt}(\d\rho^{\rm mass} + \d {\bar \rho}^{\rm mass})\sim - 2\r_{\rm th} \G_{\rm Yukawa}
\end{align}
where 
\beq
\r_{\rm th}= 0.01 \({30.25 \o g_{*s}}\)\times \bf 1
\eeq
is the yield of the thermalized left-handed leptons.
However, the deviation, $\d\rho^{\rm mass} + \d{\bar \rho}^{\rm mass}$, approaches to zero at the time scale $\D t_{llHH}(t)\simeq \G_{llHH,1}^{-1}(t)\simeq \G_{llHH,2}^{-1}(t)$. Thus the amount of deviation at time $t<t_{\rm ini}+\tl{t}_{Y}$ can be estimated by the integration of \Eq{source} over the time scale $\D t_{llHH}$,
\beq
\laq{source2}
\d\rho^{\rm mass}(t) + \d{\bar \rho}^{\rm mass}(t)\sim - \int_{t-\D t_{llHH}}^{t}{d t_1 2\r_{\rm th} \G_{\rm Yukawa}(t_1) }.
\eeq
 Taking $t_{\rm ini}=t_R$, $t_{\rm cut}=t_{\rm ini}+\tl{t}_{Y}$ and substituting \eq{source2}, one can solve \Eq{First_equation} and obtains 
\beq
{\tl{\D}^{(0)}}_{12}=(\tl{\D}^{(0)}_{21})^*\sim i \Im{\(i\int_{t_{\rm ini}}^{t_{\rm cut}}{dt_1 \int_{t_1}^{t_{\rm 1}+\D t_{llHH}}{ dt_2 [ \Omega_T^{\rm mass}(t_1), 2 \G_{\rm Yukawa}(t_2)] \r_{\rm th}}}\)_{12}},
\eeq
while other components of $\tl{\D}^{(0)}$ are nearly zero due to the wash-out effect.
Here we have used the fact that only the imaginary part of ${\tl{\D}^{(0)}}_{12}=(\tl{\D}^{(0)}_{21})^*$ conserves due to the approximate $\SO(2)$ symmetry. 
Since the second term in \Eq{thermalmass} is important for $\Omega_T^{\rm mass}$ in the commutation relation, the $t_1$ integration dominates at around $t_1 \sim t_{\rm cut}$. 
For the normal mass ordering, one obtains
\begin{align}
\laq{D12RL}
\tl{\D}^{(0)}_{1 2}&\sim  \left. -2 i \D m_{21}^2  \Re[\(\G_{\rm Yukawa}\)_{12}] \r_{\rm th}  \D{t}_{llHH} t \right|_{t=t_{\rm ini}+\tl{t}_{Y}}\\
 &\sim 4\times 10^{-8} i  \cos{\a_M\o2}\({(\D m_{21}^2)^{\rm pole}\o \(0.009\EV\)^2}\)\( {(0.1 \EV)^2 \o (\ol{m}^2_\n)^{\rm pole}}\)\cdot \({30.25 \o g_{*s}}\) .
\end{align}
By employing \Eq{asydege}, the net asymmetry is obtained as 
\begin{align}
\laq{Rlep}
{n_L\o s}\sim &\( -2\times 10^{-10} \sin{\a_M} + 10^{-10} \sin\(\a_M-\d\)+ 4\times 10^{-11} \sin\(\a_M+\d\)\) \non \\
&              ~~~~~~~~~~~~~~~~~~~~~~~~~~~~~~~~~~~~~~~~~~~~~~~~~~\times \({(\D m_{21}^2)^{\rm pole}\o \(0.009\EV\)^2}\) \cdot \({30.25 \o g_{*s}}\).
\end{align}
As indicated from the parameter dependence, the formula can also apply to the inverted mass ordering with $\ol{m}_{\n}=\O(0-0.1)\EV$. This can be seen from the fact that it fits well with the numerical results in Figs. \ref{fig:Numerical_Result_Higgsdege}, \ref{fig:deltadegedep}, and the inverted mass hierarchy cases in Figs. \ref{fig:Numerical_Result_Higgs} and \ref{fig:deltadep}.

In fact, $\tl{\D}^{\rm mass}_{13}\AND \tl{\D}^{\rm mass}_{23}$ produced at $t\sim \tl{t}_Y$ would not be destroyed with $\ol{m}_\n >\O(0.1)\EV$ (See \Eq{offdiag}).\footnote{Although, it is disfavored from the Planck data and baryon acoustic oscillation measurement \cite{Ade:2015xua}.}
The corresponding asymmetry can be calculated from the same procedure and we do not discuss this further. \\

Notice that with high reheating temperature, the asymmetry is dominantly
generated from the departure of the thermal equilibrium of the left or
right-handed leptons, which results from the decoupling of the gauge
interactions. This does not depend much on the precise information for
the inflaton decay products. In particular, the amount of the asymmetry
is independent of the $B$, $m_\f \AND T_R$. The UV insensitive amount
is, interestingly, around the order of the observed one for $\O(1)$ CP
phases in the PMNS matrix.  This indicates, by taking into account the quantum mechanics, 
a general thermalization process can lead to a good opportunity for baryogenesis. 
\section{Summary}

The neutrino oscillation has been understood as the macroscopic quantum
interference phenomena. The neutrinos traveling in the sun, atmosphere
and also terrestrial baselines are superpositions of the waves with
different frequencies and thus the probability of observing some flavor
becomes dependent on the travel distances.

In the early Universe, the whole Universe can be thought of as a
high-temperature medium. The neutrinos (and also charged leptons)
traveling through the medium undergo the flavor oscillation of the
cosmic size. Even though the neutrino masses are tiny enough to be
ignored in the high-temperature medium, the Universe is in fact opaque
and the matter effects are important for leptons/neutrinos due to
various interactions such as the gauge interactions, the lepton Yukawa
interactions as well as the lepton number violating $llHH$ interaction
if the neutrinos are Majorana particles.

At the very first stage of the Universe, the leptons are produced
through the decays of inflatons. The quantum states of these leptons can
be described by density matrices. The scattering with the medium reduces
the matrix into a diagonal form in some basis. For example, the pair
annihilation process brings the sum of the density matrices of the
leptons and anti-leptons into the one proportional to the unit matrix,
which stops the oscillation effects. Also, the scatterings through the
lepton Yukawa and the $llHH$ interactions bring the density matrices
into diagonal forms in the flavor and the mass basis, respectively.
One can think of these scattering processes as ``observations.'' Through
these observations, the density matrices evolve non-trivially and
settle into a form deviated from the thermal equilibrium due to the
cosmic expansion.

We find through the numerical analyses the lepton number is indeed
generated by these quantum effects. 
In particular, if the inflaton decays
into the Higgs boson dominantly, the high energy leptons are generated
as secondary products via the scattering through the $llHH$
interactions. In this case, the leptons are in the neutrino ``mass''
eigenstates. Since the effective Hamiltonian is ``flavor'' diagonal due
to the thermal masses from Yukawa interactions, the oscillation takes
place. The net lepton asymmetry is produced by the subsequent scattering
processes. The source of the CP violation is the Dirac and Majorana
phases in the PMNS matrix, and enough amount of asymmetry can be
produced for high enough reheating temperatures.

There is always a contribution to the baryon asymmetry of the Universe from the flavor oscillations of the leptons in the inflationary scenario. Our numerical results have shown that the successful baryogenesis is possible in any models to explain the neutrino masses by the $llHH$ terms at low energy. At least, it works if the UV scale to generate the $llHH$ terms is higher than $10^8$~GeV.


\section*{Acknowledgements} 
WY thanks the hospitality of the KEK theory group during his visit. 
WY thanks Hiroyuki Ishida for the useful discussion on the kinetic equation.  
This work is supported by JSPS KAKENHI Grant-in-Aid for Scientific
Research (No.~15H03669 and 15KK0176 [RK]), MEXT Grant-in-Aid for Scientific
Research on Innovative Areas (No.~25105011 [RK] and 18H05542 [RK]) and Grant-in-Aid for
JSPS Fellows (No.~16J06151 [YH]).

\appendix
\section{Inflaton decay}\label{app:inflaton_decay}
If the inflaton $\phi$ is gauge singlet, the decay is described by
\begin{align}
\mathcal{L}_{\rm decay}=
y m_\phi \phi H^\dagger H+\left({c_1 \over\Lambda}\phi\bar{L}HE+{\rm h.c.}\right)+{c_2 \over\Lambda} \phi F_{\mu\nu}F^{\mu\nu}+{c_3 \over\Lambda} \phi F_{\mu\nu}\tilde{F}^{\mu\nu}+...,
\end{align}
where $...$ represents other decay channel which is not relevant in the following discussion.
Unless the first term is small, the main decay channel is Higgs boson, and the reheating temperature is given by
\begin{align}\label{eq:reheating_dim4}
T_{R, {\rm dim}4}\simeq 3\times10^{13}\GEV\left(y\over10^{-2}\right)\left(m_\phi\over10^{14}\GEV\right).
\end{align}
On the other hand, if the dimension $4$ term is somehow suppressed, the decay to the gauge bosons is important, and the decay to leptons is suppressed due to the three body decay. The reheating temperature and branching fraction to the leptons are
\begin{align}
&T_{R, {\rm dim}5}\simeq 2\times10^{12}\GEV\left(m_\phi\over10^{14}\GEV\right)^{3/2}\left(10^{17}\GEV\over\Lambda\right),
&&  B\sim10^{-2},
\end{align}
assuming that $c_{1,2,3}=\mathcal{O}(1)$.

If the $\phi$ has the gauge charge same as Standard Model Higgs boson, we can write the dimension $4$ coupling
\begin{align}
\mathcal{L}\sim y'\bar{L}\phi E+ {\rm h.c.} 
\end{align}
In this case, the reheating temperature is same as Eq.~\eqref{eq:reheating_dim4} except for the replacement $y\to y'$.

Therefore, we can obtain the reheating temperature and branching fraction which realize the successful baryogenesis.

\section{Kinetic equation for right-handed leptons}\label{App:right-handed_leptons}

For completeness, here the kinetic equation including the right-handed leptons is presented although the numerical impacr is small. 
The right-handed neutrino gives rise the addition term to $\left(\d  \G_T^{p}\right)_{ij}$, which is given by 
\begin{align}
{3y_t^2 T\over32\pi^3}y_i\(-\(\d \ol{\r}_R\)^t+2\d \r_R\)_{ij}y_j
\end{align}
The kinetic equation for the leptons is
\begin{align}
  &i\frac{d \d \rho_{R}}{dt} = [\Omega_{ R} , \d \rho_{ R}] - 
  \frac{i}{2} \{ \Gamma_{ R}^d, \d \rho_{ R}  \}
  + i \Gamma_{ R}^p
  + i \Gamma_{ R}^{\rm pair},
  \\
  &\(\Omega_{ R}\)_{ij}={y_i^2\over8}T\delta_{ij},
  \quad\quad\quad
  \(\Gamma_{ R}^p\)_{ij}=
  {3y_t^2 \over64\pi^3}{T\over|{\bf k}|}
  \left[2y_i\(\d\r_T{|{\bf k}|\over T}+\r_k\)_{ij}y_j-y_i\(\d\ol{\r}_T{|{\bf k}|\over T}+\ol{\r}_k\)_{ji}y_j\right],
  \\
  &\(\Gamma_{ R}^d\)_{ij}=
  {9y_t^2\over 32\pi^3}T y_i^2\delta_{ij},
  \quad\quad\quad
  \(\Gamma_{ R}^{\rm pair}\)_{ij}=-{C_Y \over2}\alpha_Y^2\(\d \r_R+\d \ol{\r}_R\)_{ij}.
\end{align}
where $C_Y$ represents the uncertainty where we have taken to be $C_Y=C$ in the numerical calculation.
In the inflaton decay to lepton case,
\beq \quad \d \r_T|_{t=t_R} = \d \bar \r_T|_{t=t_R}= 0.
 \eeq
is added to \Eq{initial} as an initial condition. 
When the inflaton dominantly decays to Higgs bosons, the initial condition is changed to be
\beq
\quad \d \r_T|_{t=t_R} = \d \bar \r_T|_{t=t_R}= -0.002 \({100\o g_{*s}(T_R)}\).
\eeq

\section{Couplings used in numerical calculation}\label{App:parameters}
We have used the SM couplings evolved to the scale $10^{12}\GEV\text{--}10^{13}\GEV$~\cite{Xing:2011aa,Hamada:2012bp,Buttazzo:2013uya}:
\begin{align}
& g_Y= 0.42, 
&& g_2=0.55,  
&& y_t=0.47, 
&& y_{\mu}=5.8\times 10^{-4}, 
&& y_{\t}=9.8\times 10^{-3}. 
\end{align}
The $llHH$ interaction has an overall factor  \cite{Chankowski:1993tx}
\beq
m_{\nu \a }=1.27 m_{\nu \a}^{\rm pole}.\eeq
where 
the right hand side is the experimental value given in \cite{Patrignani:2016xqp}.

\bibliographystyle{TitleAndArxiv}
\bibliography{Bibliography}

\end{document}